\documentclass[a4paper, journal, comsoc]{IEEEtran}
\usepackage[left=0.75in, right=0.75in, top=0.75in, bottom=4.5cm]{geometry}  

\usepackage{cite}
\usepackage{amsmath,amssymb,amsfonts}
\usepackage{algorithmic}
\usepackage{graphicx}
\usepackage{textcomp}
\usepackage{xcolor}

\usepackage{graphicx}

\usepackage{subcaption}
\def\BibTeX{{\rm B\kern-.05em{\sc i\kern-.025em b}\kern-.08em
    T\kern-.1667em\lower.7ex\hbox{E}\kern-.125emX}}

\RequirePackage{etex}

\usepackage{babel}
\usepackage{graphicx}
\usepackage{adjustbox}
\usepackage[font=small]{caption}
\usepackage{flushend}
\usepackage[mode=tex]{standalone} 

\usepackage{amsmath,amssymb,amsfonts,bbm,bm,mathrsfs}
\usepackage{color}
\usepackage{algorithmic}
\usepackage{graphicx}
\usepackage{textcomp}
\usepackage{xcolor}
\usepackage{amsthm}

\usepackage[ruled,vlined,linesnumbered]{algorithm2e}

\usepackage[mode=tex]{standalone} 

\usepackage{tikz}

\usepackage{chronosys}
\usepackage{xspace}
\usetikzlibrary{shapes,arrows,calc,positioning,fit,automata}
\usetikzlibrary{decorations.markings}
\usetikzlibrary{overlay-beamer-styles}
\usepackage{fontawesome}

\usepackage{pgfplots}
\usepackage{pgfplotstable}
\usepackage{filecontents}

\usepackage{caption,subcaption}

\usepackage{soul}

\newcommand{\Xcal}{\mathcal{X}}

\newcommand{\E}{\mathbb{E}}

\DeclareMathOperator*{\argmax}{arg\,max}

\newcommand{\myBlue}{blue!80!black}
\newcommand{\myGreen}{green!60!black}
\newcommand{\myRed}{red!80!black}

\newcommand{\Enc}{\mathrm{e}_\theta}
\newcommand{\Demod}{\mathrm{p}_\phi}
\newcommand{\Demodprime}{\mathrm{p}_\xi}

\newcommand{\EntMod}{\mathrm{q}_\zeta}

\definecolor{NYUviolet}{HTML}{57068c} 	
\definecolor{NYUlight}{HTML}{8900e1} 	
\definecolor{NYUdark}{HTML}{330662} 	
\definecolor{NYUnight}{HTML}{220337} 	

\tikzstyle{block}=[rectangle,draw,very thick,fill=white,align=center,font=\Large]
\tikzstyle{edge} = [draw,very thick,->,-triangle 45]
\tikzstyle{plateBlue} = [draw=\myBlue, shape=rectangle, rounded corners=0.5ex, ultra thick,
minimum width=3.1cm, text=\myBlue, text width=3.1cm, align=right, inner sep=10pt, inner ysep=10pt,append after command={node[font=\bf,text=\myBlue,above left= 3pt of \tikzlastnode.south east] {#1}}]

\tikzstyle{plateRed} = [draw=\myRed, shape=rectangle, rounded corners=0.5ex, ultra thick,
minimum width=3.1cm, text=\myRed, text width=3.1cm, align=right, inner sep=10pt, inner ysep=10pt,append after command={node[font=\bf\LARGE,text=\myRed,above= 3pt of \tikzlastnode.south] {#1}}]

\tikzstyle{plateGreen} = [draw=\myGreen, shape=rectangle, rounded corners=0.5ex, ultra thick,
minimum width=3.1cm, text=green!80!black, text width=3.1cm, align=right, inner sep=10pt, inner ysep=10pt,append after command={node[font=\bf\LARGE,text=\myGreen,above= 3pt of \tikzlastnode.south] {#1}}]

\tikzstyle{plate} = [draw, shape=rectangle, rounded corners=0.5ex, ultra thick,
minimum width=3.1cm,  text width=3.1cm, align=right, inner sep=10pt, inner ysep=10pt,append after command={node[font=\bf\Large,above left= 3pt of \tikzlastnode.south east] {#1}}]

\tikzstyle{plateSmall} = [draw, shape=rectangle, rounded corners=0.5ex, ultra thick,
minimum width=3.1cm,  text width=3.1cm, align=right, inner sep=10pt, inner ysep=10pt,append after command={node[font=\bf,above= 3pt of \tikzlastnode.south] {#1}}]

\tikzstyle{plateUp} = [draw, shape=rectangle, rounded corners=0.5ex, ultra thick,
minimum width=3.1cm,  text width=3.1cm, align=right, inner sep=10pt, inner ysep=10pt,append after command={node[font=\bf\Large,above= 3pt of \tikzlastnode.north] {#1}}]

\tikzstyle{plateUpEast} = [draw, shape=rectangle, rounded corners=0.5ex, ultra thick,
minimum width=3.1cm,  text width=3.1cm, align=right, inner sep=10pt, inner ysep=10pt,append after command={node[font=\bf\Large,above= 3pt of \tikzlastnode.north east, anchor=south east] {#1}}]

\tikzstyle{plateDown} = [draw, shape=rectangle, rounded corners=0.5ex, ultra thick,
minimum width=3.1cm,  text width=3.1cm, align=right, inner sep=10pt, inner ysep=10pt,append after command={node[font=\bf\Large,below= 3pt of \tikzlastnode.south] {#1}}]

\usepackage[prefix=]{xcolor-material}
\pgfmathdeclarerandomlist{material}{%
	{Red}{Blue}{Green}}
\tikzset{%
	half clip/.code={
		\clip (0, -256) rectangle (256, 256);
	},
	color/.code=\colorlet{fill color}{#1},
	color alias/.code args={#1 as #2}{\colorlet{#1}{#2}},
	colors alias/.style={color alias/.list/.expanded={#1}},
	execute/.code={#1},
	on left/.style={.. on left/.style={#1}},
	on right/.style={.. on right/.style={#1}},
	split/.style args={#1 and #2}{
		on left ={color alias=fill color as #1},
		on right={color alias=fill color as #2, half clip}
	}
}
\newcommand\reflect[2][]{%
	\begin{scope}[#1]\foreach \side in {-1, 1}{\begin{scope}
				\ifnum\side=-1 \tikzset{.. on left/.try}\else\tikzset{.. on right/.try}\fi
				\begin{scope}[xscale=\side]#2\end{scope}
\end{scope}}\end{scope}}

\tikzset{%
	cat/.pic={
		\tikzset{x=1.5cm/5,y=1.5cm/5,shift={(0,-1/3)}}
		\useasboundingbox (-1,-1) (1,2);
		\fill [BlueGrey900] (0,-2)
		.. controls ++(180:3) and ++(0:5/4) .. (-2,0)
		arc (270:90:1/5)
		.. controls ++(0:2) and ++(180:11/4) .. (0,-2+2/5);
		\foreach \i in {-1,1}
		\scoped[shift={(1/2*\i,9/4)}, rotate=45*\i]{
			\clip [overlay] (0, 5/9) ellipse [radius=8/9];
			\clip [overlay] (0,-5/9) ellipse [radius=8/9];
			\fill [BlueGrey900] ellipse [radius=1];
			\clip [overlay] (0, 7/9) ellipse [radius=10/11];
			\clip [overlay] (0,-7/9) ellipse [radius=10/11];
			\fill [Purple100] ellipse [radius=1];
		};
		\fill [BlueGrey900] ellipse [x radius=3/4, y radius=2];
		\fill [BlueGrey100] ellipse [x radius=1/3, y radius=1];
		\fill [BlueGrey900]
		(0,15/8) ellipse [x radius=1, y radius=5/6]
		(0, 8/6) ellipse [x radius=1/2, y radius=1/2]
		{[shift={(-1/2,-2)}, rotate= 10]  ellipse [x radius=1/3, y radius=5/4]}
		{[shift={( 1/2,-2)}, rotate=-10] ellipse [x radius=1/3, y radius=5/4]};
		\fill [BlueGrey500]
		(-1/9,11/8) ellipse [x radius=1/5, y radius=1/5]
		( 1/9,11/8) ellipse [x radius=1/5, y radius=1/5];
		\fill [Purple100]
		(0,12/8)     ellipse [x radius=1/10, y radius=1/5]
		(0,12/8+1/9) ellipse [x radius=1/5 , y radius=1/10];
		\foreach \i in {-1,1}
		\scoped[shift={(1/2*\i,2)}, rotate=35*\i]{
			\clip [overlay] (0, 1/7) ellipse [radius=2/7];
			\clip [overlay] (0,-1/7) ellipse [radius=2/7];
			\fill [Yellow50] ellipse [radius=1];
		};
		\scoped{
			\clip (-1,-2) rectangle ++(2,1);
			\fill [BlueGrey900] (0,-2) ellipse [radius=1/2];
			\fill [Grey100]
			(-1/2,-2) ellipse [x radius=1/3, y radius=1/4]
			( 1/2,-2) ellipse [x radius=1/3, y radius=1/4];
		};
		\foreach \i in {-1,1}
		\foreach \j in {-1,0,1}
		\fill [Grey100, shift={(0,11/8)}, xscale=\i, rotate=\j*15,
		shift=(0:1/2)]
		ellipse [x radius=1/3, y radius=1/64];
	},
dog/.pic={
	\begin{scope}[x=1.5cm/480,y=1.5cm/480]
		\useasboundingbox (-256, -256) (256, 256);
		\reflect[split=Brown400 and Brown500]{
			\fill [fill color] (0,-64) ellipse [x radius=160, y radius=144];
			\fill [fill color] (0, 32) ellipse [x radius=128, y radius=112];
			\fill [fill color] (32,96)
			.. controls ++( 75:128) and ++(105:128) .. ++(192,  0)
			.. controls ++(285: 96) and ++(285: 96) .. ++(-80,-32)
			.. controls ++(105: 64) and ++( 75: 32) .. cycle;
		}
		\reflect[split={Grey100 and Grey200}]{
			\clip (0,-64) ellipse [x radius=160, y radius=144];
			\fill [fill color](0,-224) 
			.. controls ++(  0:64) and ++(270:64) .. ++(96,64)
			.. controls ++( 90:64) and ++(270:64) .. ++(-96,192)
			.. controls ++(270:64) and ++( 90:64) .. ++(-96,-192)
			.. controls ++(270:64) and ++(180:64) .. cycle;
		}
		\reflect[split={Pink100 and Pink200}]{
			\fill [fill color](0,-192) ellipse [x radius=28, y radius=32];
		}
		\reflect[split={BlueGrey800 and BlueGrey900}]{
			\fill [fill color](0,-144) 
			.. controls ++(  0:20) and ++(315:20) .. ++( 40,64)
			.. controls ++(135:20) and ++( 45:20) .. ++(-80, 0)
			.. controls ++(225:20) and ++(180:20) .. cycle;
			\fill [BlueGrey900] (56, 0) circle [radius=20];
			\fill [fill color] (-8,-112)
			-- ++(16,0) -- ++(0,-32) arc (180:360:24)
			arc (180:0:8) arc (360:180:40);
		}
\end{scope}}
}

\tikzset{
	o/.style={
		shorten >=#1,
		decoration={
			markings,
			mark={
				at position 1
				with {
					\draw circle [radius=#1];
				}
			}
		},
		postaction=decorate
	},
	o/.default=2pt
}

\tikzset{naming/.style={align=center,font=\large}}
\tikzset{antenna/.style={insert path={-- coordinate (ant#1) ++(0,0.25) -- +(135:0.25) + (0,0) -- +(45:0.25)}}}
\tikzset{station/.style={naming,draw,shape=dart,shape border rotate=90, minimum width=15mm, minimum height=30mm,outer sep=0pt,inner sep=3pt}}
\tikzset{stationPoster/.style={naming,draw,shape=dart,shape border rotate=90, minimum width=20mm, minimum height=40mm,outer sep=0pt,inner sep=3pt}}
\tikzset{mobile/.style={naming,draw,shape=rectangle,minimum width=12mm,minimum height=6mm, outer sep=0pt,inner sep=3pt}}
\tikzset{radiation/.style={{decorate,decoration={expanding waves,angle=90,segment length=4pt}}}}

\begin{document}

\title{
    Learning-Based Compress-and-Forward Schemes for the Relay Channel
}

\author{
    Ezgi~Ozyilkan$^*$,~\IEEEmembership{Graduate~Student~Member,~IEEE},
    Fabrizio~Carpi$^*$,~\IEEEmembership{Graduate~Student~Member,~IEEE},  
    Siddharth~Garg,~\IEEEmembership{Member,~IEEE},
    Elza~Erkip,~\IEEEmembership{Fellow,~IEEE,}%
        \thanks{A preliminary version of this work was presented at the IEEE Workshop on Signal Processing Advances in Wireless Communications, 2024~\cite{ozyilkan2024relay}.
        The work of E.~Ozyilkan, F.~Carpi, S.~Garg and E.~Erkip was supported in part by the NYU WIRELESS Industrial Affiliates Program, and by the NSF grants \#1925079 and \#2003182. \emph{(Corresponding authors: Ezgi Ozyilkan and Fabrizio Carpi.)}
    }%
    \thanks{ 
        E.~Ozyilkan, F.~Carpi, S.~Garg and E.~Erkip are with the Department of Electrical and Computer Engineering, New York University, Brooklyn, NY (emails: \texttt{\{ezgi.ozyilkan, fabrizio.carpi, siddharth.garg, elza\}@nyu.edu}). 
    }%
    \thanks{
        $^*$Equal contribution.
    }%
}%

\thispagestyle{empty}
\maketitle
\thispagestyle{empty}

\begin{abstract}
The relay channel, consisting of a source-destination pair along with a relay, is a fundamental component of cooperative communications. 
While the capacity of a general relay channel remains unknown, various relaying strategies, including compress-and-forward (CF), have been proposed. In CF, the relay forwards a quantized version of its received signal to the destination. Given the correlated signals at the relay and destination, distributed compression techniques, such as Wyner--Ziv coding, can be harnessed to utilize the relay-to-destination link more efficiently. 
Leveraging recent advances in neural network-based distributed compression, 
we revisit the relay channel problem and integrate a learned task-aware Wyner--Ziv compressor into a primitive relay channel with a finite-capacity out-of-band relay-to-destination link. The resulting neural CF scheme demonstrates that our compressor recovers binning of the quantized indices at the relay, mimicking the optimal asymptotic CF strategy, although no structure exploiting the knowledge of source statistics was imposed into the design. The proposed neural CF, employing finite order modulation, operates closely to the rate achievable in a primitive relay channel with a Gaussian codebook. We showcase the advantages of exploiting the correlated destination signal for relay compression through various neural CF architectures that involve end-to-end training of the compressor and the demodulator components. Our learned task-oriented compressors provide the first proof-of-concept work toward interpretable and practical neural CF relaying schemes.
\end{abstract}


\section{Introduction}
\IEEEPARstart{T}{he} relay channel, as introduced by van der Meulen~\cite{relay_initial}, is a building block of multi-user communications. 
In this model, a relay facilitates communication between a source and a destination by forwarding its ``overheard'' received signal to the destination. 
As such, the relay channel comprises a \emph{broadcast channel}, from the source to both the relay and the destination, and a \emph{multiple access channel}, from both the source and the relay to the destination. 
The relay channel forms the foundation of cooperative networking, which has been shown to be effective in mitigating fading~\cite{sendos,laneman}, increasing data rates~\cite{DF_1}, and managing interference~\cite{gesbert1}. 
With the advent of 6G, new forms of relaying and cooperation are envisioned for communicating in highly dynamic settings~\cite{gesbert2, sidelink-5G}.

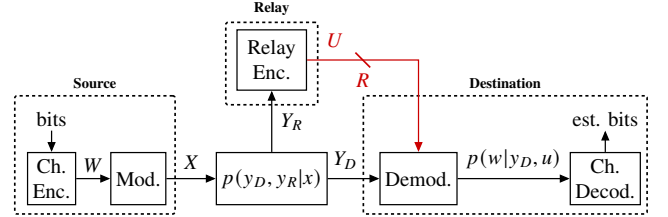
\begin{figure}[t]
    \centering
    \begin{tikzpicture}
\tikzstyle{block}=[rectangle,draw,very thick,fill=white,align=center,font=\huge]
    \def\dhor{4.5}
    \def\dvert{4}

    \node[block, draw=white] (S) at (-0.65*\dhor,0.5*\dvert) {bits};
    
    \node[block, inner sep=5pt, align=center, minimum height = 50] (chenc) at (-0.65*\dhor,0) {Ch.\\Enc.};
    \node[block, inner sep=5pt, align=center, minimum height = 50] (mod) at (0,0) {Mod.};
    \node[plateUp={Source}, draw=black, dashed, fit=(chenc)(mod)(S), inner sep=10pt] () at(-0.5*0.65*\dhor,0.2*\dvert)  {};
    
    \node[block, inner sep=5pt, align=center, minimum height = 50] (channel) at (\dhor,0) {$p(y_D, y_R \vert x)$};

    \node[block, inner sep=10pt, align=center] (relay) at (\dhor,\dvert) {Relay\\Enc.};
    \node[plateUp={Relay}, draw=black, dashed, fit=(relay), inner sep=10pt] () at(\dhor,0.97*\dvert)  {};

    \node[block, inner sep=5pt, align=center, minimum height = 50] (dec) at (2.1*\dhor,0) {Demod.};
    \node[block, inner sep=5pt, align=center, minimum height = 50] (chdec) at (3.5*\dhor,0) {Ch.\\Decod.};
    \node[block, draw=white] (Shat) at (3.5*\dhor,0.5*\dvert) {est. bits};
    \node[plateUp={Destination}, draw=black, dashed, fit=(chdec)(dec)(Shat), inner sep=10pt] () at(2.75*\dhor,0.2*\dvert)  {};

    \path[edge] (S) -- (chenc);
    \path[edge] (chenc) --node[block, draw=white, inner sep=0, above=0.2]{$W$} (mod);
    \path[edge] (mod) --node[block, draw=white, above=0.2]{$X$} (channel);
    \path[edge] (channel) --node[block, draw=white, pos=0.35, above=0.2]{$Y_D$} (dec);
    \path[edge] (channel) --node[block, draw=white, right=0.2]{$Y_R$} (relay);
    \path[edge] (dec) --node[block, draw=white, above=0.2]{$p(w \vert y_D,u)$} (chdec);
    \path[edge] (chdec) -- (Shat);
    \path[edge, \myRed] (relay) -- node[block, draw=white, pos=0.25, above=0.3]{$U$} node[block, draw=white, below=0.3]{$R$} node {\tikz \draw[ultra thick, solid,-] (0,0) -- +(0.5,-0.5);} (2.1*\dhor,\dvert) -- (dec);

\end{tikzpicture}
    \caption{
        The \emph{primitive} relay channel (PRC) under consideration. 
        The red link denotes out-of-band relaying between the relay and the destination.
    }
    \label{fig:sys_model}
\end{figure}

Despite decades of research, the capacity of the general relay channel is still unknown to this day.
Cover and El~Gamal~\cite{relaycapacity} provided upper and lower bounds for the general relay channel by invoking information theoretic achievability and converse arguments. 
These bounds coincide only in a few special cases, such as the physically degraded Gaussian relay channel. 
Even though optimum relaying strategies are not known in general, various effective relaying techniques have been proposed, which can be broadly categorized into two main classes: \emph{decode-and-forward} (DF) and \emph{compress-and-forward} (CF); see~\cite{relaycapacity} for a detailed analysis of DF, CF, their variations and combinations. 
While DF is known to be efficient in certain scenarios~\cite{DF_1}, its achievable rate is bounded by the capacity of the source-to-relay channel since the relay is required to perfectly decode the source information. 

On the other hand, in CF,  the relay refrains from directly decoding the source and instead, compresses its received signal to send to the destination. 
Upon reception of the compression index, the destination combines it with its own received signal to decode the source information.  
Given that the received signals at the relay and destination are correlated, the relay can leverage \emph{distributed compression} techniques to reduce the compression rate without requiring explicit knowledge of the received signal at the destination.
As such, it can utilize Wyner--Ziv (WZ) source coding~\cite{Wyner_Ziv}, also known as source coding with decoder-only side information, to efficiently describe its received signal. 
Unlike DF, CF relaying consistently outperforms direct transmission since the relay always aids in communication, even when the source-to-relay channel is poor. 
For additional discussion on scenarios where CF has been proven to be optimal, we direct readers to~\cite{kang2008capacity}.  
Despite its benefits, the limitations of practical WZ implementations operating in the finite blocklength regime have hampered the widespread use of CF relaying.

In this paper, drawing on recent advances in neural distributed compression~\cite{Ezgi-ISIT-2023, ozyilkan2024distributed}, we revisit practical CF design and illustrate the potentials of learning for reaping the benefits of CF. To highlight design constraints for CF, we focus on the \emph{primitive} relay channel (PRC)~\cite{Kim2008CodingTF}, depicted in Fig.~\ref{fig:sys_model}, where there is an orthogonal (out-of-band) noiseless link of rate $R$ connecting the relay to the destination. Our main contributions are summarized as follows:

\begin{itemize}
    \item We present learned CF relaying schemes for the Gaussian PRC that are based on 
    task-aware neural distributed compressors, where the task is to maximize the source-to-destination communication rate. We provide several architectures, differing in the way the distributed compression is carried out. Each of these schemes consists of a compressor at the relay and (soft) demodulator at the destination, both of which are learned in an end-to-end fashion. 
    \item  We offer post-hoc interpretations of the resulting neural CF schemes on some representative modulation schemes. Building on the findings presented in~\cite{ozyilkan2024relay}, we provide an in-depth analysis demonstrating that the task-aware neural relay quantizer exhibits \emph{binning} (grouping) in the source space, which is known to be information theoretically optimal.
    In addition, we illustrate explainable decision boundaries for the learned demodulator at the destination. These structures emerge from learning, not from design choices based on system parameters.
    \item Using a comprehensive set of experimental results, we evaluate the performance of our neural CF strategies both in terms of communication and error rates. Beyond the evaluation presented in~\cite{ozyilkan2024relay}, we extend our analysis to include higher-order modulation schemes, such as 8-PAM, as well as complex-valued ones, including 4-QAM and 16-QAM. Comparison with theoretical benchmarks suggests the effectiveness of our learning-based relaying frameworks.   
    \item We provide a detailed analysis of robustness to varying signal-to-noise ratios (SNRs) both at the relay and the destination. We empirically demonstrate that training over a range of SNRs enables the resulting CF strategy to maintain good performance across the range of interest.
\end{itemize}

Overall, our learned CF framework represents the first proof-of-concept investigation towards practical and robust CF relaying, with the added benefit of yielding interpretable results. 

A few comments are in order regarding our motivation for considering the PRC. Firstly, the PRC offers a scenario where the compressed relay signal can be readily transmitted to the destination. 
Compared to the general relay channel, the PRC model decouples the relay transmission from that of the source, allowing a natural setting to study CF. 
Note that the PRC model represents the simplest channel coding problem, viewed from the source's perspective, with a rate constraint among the two receiving terminals (relay and destination). Simultaneously, it also encapsulates the simplest compression problem, viewed from the relay's perspective, for enabling channel coding between the source and the destination.
Secondly, PRC provides a good model for scenarios in which a different wireless or wired interface is used for relaying, such as base station cooperation. Finally, the relaying strategies developed for the PRC can be extended to a more general relay channel model by incorporating the multi-access reception at the destination. We also note that  CF relaying is optimal for the PRC if the relay is unaware of the source codebook, also known as \emph{oblivious} relaying~\cite{Simeone_2}. The oblivious setting is well-suited to the learning framework, in which the relay is not explicitly informed about the transmission strategy used by the source. Rather, a data-driven relay trains its compressor based on samples of its channel output. 

There is limited literature addressing practical CF designs, e.g.,~\cite{nested_relay_quantizer, practical_relay_quantizer}.
Both of these works proposed entropy-constrained scalar quantizer designs with binary phase shift keying (BPSK) modulation for the half-duplex Gaussian relay channel, with~\cite{nested_relay_quantizer} considering lossless Slepian--Wolf (SW) coded nested quantization as a practical form of WZ compression (following the WZ compressor proposed in~\cite{SWCNSQ}), and~\cite{practical_relay_quantizer} not taking into account the side information at the destination while quantizing at the relay. 
In addition, these works relied on handcrafted and analytical solutions, thereby constraining their generalization to more complex communication settings. Unlike some of the previous distributed compression work (e.g.,~\cite{DISCUS, SWCNSQ}) or the aforementioned relay quantizer designs, our proposed CF strategies neither enforce any specific structure onto the model nor assume prior knowledge about the source-to-destination communication strategies or link qualities.

Recent learning approaches for the relay channel~\cite{bian2022deep, bian2024processandforward, arda2024semantic} considered a joint source-channel setting, where the first two focused on image transmission via joint source-channel coding, while the last one targeted text communication utilizing attention-based transformer architectures. 
Our paper, in contrast, concentrates {\em only on the channel} part and addresses an important open problem in the cooperative communications literature, namely how to make CF practical. While our learned CF framework is built upon those of~\cite{Ezgi-ISIT-2023, ozyilkan2023workshop, ozyilkan2023neural}, an important distinction is that in CF, the goal is to facilitate source-to-destination communication, and not to reconstruct the relay signal per se. In fact, it is demonstrated in~\cite{practical_relay_quantizer} that relay compression that minimizes mean squared error distortion can be significantly suboptimal. We refer the reader to~\cite{ozyilkan2024distributed} for an overview of distributed compression and practical designs, including those based on neural networks, that focus on signal reconstruction.

This paper is organized as follows.
The system model is explained in Sec.~\ref{sec:system_model}.
The proposed neural CF schemes and learning procedures are described in Sec.~\ref{sec:neural_schemes}. 
Extensive numerical results are presented in Sec.~\ref{sec:discussion}.
Conclusion and future work are discussed in Sec.~\ref{sec:conclusion}.

\section{System Model} \label{sec:system_model}
In this section, we introduce the PRC model (Fig.~\ref{fig:sys_model}) and provide an achievable rate for  CF, which is tight for oblivious relaying. Next, we explain the performance criterion we adopt for testing relaying schemes that involve task-aware neural distributed compressors.

\subsection{Primitive Relay Channel (PRC)}
We consider the PRC setup~\cite{Kim2008CodingTF}, illustrated in Fig.~\ref{fig:sys_model}. 
The Gaussian PRC, which we study in this paper, is given by:
\begin{align}
\begin{split}
    Y_R = h_R\; X + N_R, \\
    Y_D = h_D\; X + N_D,
\label{eq:prc}
\end{split}
\end{align}
where $X$  denotes the signal transmitted by the source, $Y_R$ and $Y_D$ denote the received signals at the relay and the destination, and $h_R$ and $h_D$ are the corresponding channel gains, respectively. 
The noise components, $N_R$ and $N_D$, are independent of one another and of $X$. 

In this work, we consider both real and complex-valued channels. 
For the real-valued channel, without loss of generality, we consider  $X, h_R, h_D\in\mathbb{R}$, $N_R\sim\mathcal{N}(0,1)$ and $N_D\sim\mathcal{N}(0,1)$.
For the complex-valued channel, we assume $X, h_R, h_D\in\mathbb{C}$, $N_R\sim\mathcal{CN}(0,1)$ and $N_D\sim\mathcal{CN}(0,1)$.
Note that by allowing for arbitrary $(h_R, h_D)$, one can incorporate the effect of different SNRs for the source-to-relay and source-to-destination links. 
As customary, we consider communication over a blocklength of $n$, with $n$ asymptotically large, and i.i.d. noise. 
For brevity, we omit the time index in (\ref{eq:prc}). 
The out-of-band relay-to-destination channel is represented by a link with \emph{relay rate} $R$ bits/channel use.

For a general PRC $p(y_D, y_R\vert x)$ with an oblivious relay, where the relay is agnostic to the codebook shared by source and destination, it was shown that the capacity can be attained by the CF strategy with time sharing~\cite{Simeone_2}. 
Without time-sharing, the following rate $C$ is achievable~\cite{Simeone_2}:
\begin{align}
   C &= \max \;  \mathrm{I}(X;Y_D, U),
   \label{eq:opt1}\\
   \text{s.t. } R &\geq \mathrm{I}(Y_{R}; \, U \;  \vert \;  Y_{D}), 
   \label{eq:opt2}
\end{align} where maximization is with respect to the distribution $p(x)p\left(u\vert y_{R}\right)$. Here, $U$ corresponds to the relay's compressed description of $Y_R$, and the rate constraint in~\eqref{eq:opt2} coincides with the one that emerges in WZ rate--distortion function~\cite{Wyner_Ziv}.
Recall that in CF, the relay regards its received signal $Y_R$ as an unstructured random process jointly distributed with the signal received at the destination $Y_D$. 
This enables the relay to exploit  WZ compression~\cite{Wyner_Ziv}, to efficiently describe its received signal. 
We note that the capacity of the PRC without oblivious relaying constraint is still not fully characterized~\cite{Simeone_2}. 

For the real-valued Gaussian PRC in~\eqref{eq:prc}, the following CF rate is achieved with Gaussian input under power constraint $\E[\vert X\vert^2]\leq P$~\cite{Simeone_2}:
\begin{align}
\label{eq:C_CF-Simeone}
     C_\text{CF} = \frac{1}{2} \log_2 \left(
            1 + \gamma_D + \frac{\gamma_R}{1 + \frac{1+\gamma_D+\gamma_R}{(2^{2R}-1)(\gamma_D+1)}}
            \right),
\end{align} 
where $\gamma_D = \vert h_D\vert^2 P$ and $\gamma_R = \vert h_R\vert^2 P$ are SNRs at the destination and at the relay, respectively.
Note that, in the case of a complex-valued PRC, the factor of $1/2$ in~\eqref{eq:C_CF-Simeone} is removed.
It is shown in~\cite{Simeone_2} that while the Gaussian input is not necessarily optimal, the rate in (\ref{eq:C_CF-Simeone}) is at most
$1/2$ bit away from the capacity of the Gaussian PRC, even if the relay is not oblivious. Hence,
we will use~\eqref{eq:C_CF-Simeone} as a benchmark for our learned CF communication rates. 

\subsection{Performance Criterion}
\label{sec:perf}
For our learning-based CF frameworks, we assume a finite order modulation such that an index $W\in\{1,\dots,\vert\mathcal{X}\vert\}$, which represents the output of the channel encoder, is mapped to a symbol $X\in\mathcal{X}$, where $\mathcal{X}\subset\mathbb{R}$ (or, $\mathcal{X}\subset\mathbb{C}$ for complex-valued signals) is a constellation of cardinality $\vert\mathcal{X}\vert$. 
We consider a fixed modulation scheme with equally likely symbols, and do not optimize over the constellation $\mathcal{X}$ or over the distribution $p(x)$. 
Incorporating the learned probabilistic and geometric constellation shaping~\cite{Stark-2019} into our neural CF frameworks is beyond the scope of this work.    
Our goal is to jointly learn the {\em encoder} at the relay, which outputs a compressed description $U$,  and the (soft) {\em demodulator} at the destination, which outputs a probability distribution on $W$ (Fig.~\ref{fig:sys_model}) that maximize the mutual information $\mathrm{I}(X;Y_D, U)$ subject to the relay rate constraint $R$, as in~\eqref{eq:opt1} and~\eqref{eq:opt2}.
We assume the availability of good channel codes to be used in conjunction with the modulation scheme, and as such the mutual information $\mathrm{I}(X;Y_D, U)$ can be viewed as a CF achievable rate.  
In Sec.~\ref{sec:loss}, we will discuss how this performance criterion is incorporated into the objective function used in the learning process.

\section{Neural Compress-And-Forward (CF) Schemes} \label{sec:neural_schemes}

\begin{figure}
    \centering
    \begin{subfigure}[b]{\columnwidth}
      \begin{tikzpicture}
    \def\dhor{3}
    \def\dvert{2}
    \def\dR{0.5}

    \node[block, font=\LARGE, draw=white] (Yr) at (-0.75*\dhor,0) {$Y_R$};
    \node[block, font=\LARGE, inner sep=5pt, align=center, minimum height = 50] (enc) at (0,0) {Enc.\\$\color{\myBlue}\Enc$};
    \node[block, font=\LARGE, inner sep=5pt, align=center, minimum height = 50] (enc2) at (\dhor,0) {Entropy\\Enc.};
    \node[block, font=\LARGE, inner sep=5pt, align=center, minimum height = 50] (dec2) at (2.5*\dhor,0) {Entropy\\Dec.};
    \node[block, font=\LARGE, inner sep=5pt, align=center, minimum height = 50] (dec) at (3.5*\dhor,0) {Demod.\\$\color{\myBlue}\Demod$};
    \node[block, font=\LARGE, draw=white, anchor=west] (out) at (4.15*\dhor,0) {$\Demod(w|y_D,\Enc(y_R))$};
    \node[block, font=\LARGE, align=center, draw=white] (Yd) at (3.5*\dhor,-\dvert) {$Y_D$};

    \path[edge] (Yr) -- (enc);
    \path[edge] (enc) -- (enc2);
    \path[edge, \myRed] (enc2) -- node[block, font=\LARGE, draw=white, below=0.5]{$R$} node[block, font=\LARGE, draw=white, above=0.5]{$\color{\myBlue}\EntMod$} node{\tikz \draw[ultra thick, solid,-] (-\dR,-\dR) -- +(\dR,\dR);} (dec2);
    \path[edge] (dec2) -- (dec);
    \path[edge] (dec) -- (out);
    \path[edge] (Yd) -- (dec);

    \draw [decorate,decoration={brace,amplitude=20pt}, ultra thick]
(-1, 1.1) -- (\dhor+1.2,1.1) node [block, font=\LARGE, draw=none, above=0.8, midway] {Relay};
    \draw [decorate,decoration={brace,amplitude=20pt}, ultra thick]
(-1.2+2.5*\dhor, 1.1) -- (3.5*\dhor+1.5,1.1) node [block, font=\LARGE, draw=none, above=0.8, midway] {Destination};

\end{tikzpicture}
       \caption{}
        \label{fig:marg_model} 
    \end{subfigure}
    \begin{subfigure}[b]{\columnwidth}
      \begin{tikzpicture}
    \def\dhor{3}
    \def\dvert{2}
    \def\dR{0.5}

    \node[block, font=\LARGE, draw=white] (Yr) at (-0.75*\dhor,0) {$Y_R$};
    \node[block, font=\LARGE, inner sep=5pt, align=center, minimum height = 50] (enc) at (0,0) {Enc.\\$\color{\myBlue}\Enc$};
    \node[block, font=\LARGE, inner sep=5pt, align=center, minimum height = 50] (enc2) at (\dhor,0) {SW\\Enc.};
    \node[block, font=\LARGE, inner sep=5pt, align=center, minimum height = 50] (dec2) at (2.5*\dhor,0) {SW\\Dec.};
    \node[block, font=\LARGE, inner sep=5pt, align=center, minimum height = 50] (dec) at (3.5*\dhor,0) {Demod.\\$\color{\myBlue}\Demod$};
    \node[block, font=\LARGE, draw=white, anchor=west] (out) at (4.15*\dhor,0) {$\Demod(w|y_D,\Enc(y_R))$};
    \node[block, font=\LARGE, align=center, draw=white] (Yd) at (3*\dhor,-\dvert) {$Y_D$};

    \path[edge] (Yr) -- (enc);
    \path[edge] (enc) -- (enc2);
    \path[edge, \myRed] (enc2) -- node[block, font=\LARGE, draw=white, below=0.5]{$R$} node[block, font=\LARGE, draw=white, above=0.5]{$\color{\myBlue}\EntMod$} node{\tikz \draw[ultra thick, solid,-] (-\dR,-\dR) -- +(\dR,\dR);} (dec2);
    \path[edge] (dec2) -- (dec);
    \path[edge] (dec) -- (out);
    \path[edge] (Yd) -- (3*\dhor,-0.7*\dvert) -| (dec);
    \path[edge] (Yd) -- (3*\dhor,-0.7*\dvert) -| (dec2);

\end{tikzpicture}
       \caption{}
        \label{fig:cond_model} 
    \end{subfigure}
    \begin{subfigure}[b]{\columnwidth}
      \begin{tikzpicture}
    \def\dhor{3}
    \def\dvert{3.3}
    \def\dR{0.5}

    \node[block, draw=white, font=\LARGE] (Yr) at (-0.75*\dhor,0) {$Y_R$};
    \node[block, inner sep=5pt, align=center, minimum height = 50, font=\LARGE] (enc) at (0,0) {Enc.\\$\color{\myBlue}\Enc$};
    \node[block, inner sep=5pt, align=center, minimum height = 50, font=\LARGE] (enc2) at (\dhor,0) {Entropy\\Enc.};
    \node[block, inner sep=5pt, align=center, minimum height = 50, font=\LARGE] (dec2) at (2.5*\dhor,0) {Entropy\\Dec.};
    \node[block, inner sep=5pt, align=center, minimum height = 50, font=\LARGE] (dec) at (3.5*\dhor,0) {Demod.\\$\color{\myBlue}\Demodprime$};
    \node[block, inner sep=5pt, align=center, minimum height = 50, font=\LARGE] (dec_extra) at (3.5*\dhor,-\dvert) {Demod.\\$\color{\myBlue}\Demod$};
    \node[block, draw=white, anchor=west, font=\LARGE] (out) at (4.15*\dhor,0) {$\Demodprime(w|\Enc(y_R))$};
    \node[block, draw=white, anchor=west, font=\LARGE] (out_extra) at (4.15*\dhor,-\dvert) {$\Demod(w|y_D, \Enc(y_R))$};
    \node[block, align=center, draw=white, font=\LARGE] (Yd) at (3.5*\dhor,-1.6*\dvert) {$Y_D$};

    \node[plateUp={\color{\myGreen}Pre-trained}, draw=\myGreen, font=\LARGE, dashed, fit=(enc)(enc2)(dec2)(dec), inner sep=10pt] () at (1.75*\dhor,0)  {};

    \node[plateUp={\color{orange}Fine-tuned}, draw=orange, font=\LARGE, dashed, fit=(dec_extra), inner sep=10pt] () at(3.45*\dhor+0.1,-1.05*\dvert)   {};

    \path[edge] (Yr) -- (enc);
    \path[edge] (enc) -- (enc2);
    \path[edge, draw=orange] (2.85*\dhor+0.25,0) |- (dec_extra);
    \path[edge, \myRed] (enc2) -- node[block, draw=white, below=0.5, font=\LARGE]{$R$} node[block, draw=white, font=\LARGE, above=0.5]{$\color{\myBlue}\EntMod$} node{\tikz \draw[ultra thick, solid,-] (-\dR,-\dR) -- +(\dR,\dR);} (dec2);
    \path[edge] (dec2) -- (dec);
    \path[edge] (dec) -- (out);
    \path[edge] (dec_extra) -- (out_extra);
    \path[edge] (Yd) -- (dec_extra);

\end{tikzpicture}
       \caption{}
        \label{fig:p2p_model} 
    \end{subfigure}
    \caption{
        The three proposed  neural CF schemes: (a) and (b) are based on marginal (\emph{marg.}) and conditional (\emph{cond.}) formulations, (coupled with classic either entropy or Slepian-Wolf (SW) coder) 
        respectively; (c) is the point-to-point (\emph{p2p}) scheme. 
        The learned parameters are indicated in blue.
        Note that the schemes in (a) and (b) operationally correspond to task-aware neural Wyner--Ziv compressors, since the encoder can exploit the side information $Y_D$ at the receiver side. 
        In (c), neither parameters of $\Enc$ and $\EntMod$ are updated during the fine-tuning step (only $\Demod$ is learned). In the \emph{split I-Q} variants of each scheme (not depicted), we have two separate encoders that compress in-phase and quadrature components of the complex-valued signal independently. Wherever we present relevant experiments, we label the depicted respective scheme illustrated in this figure as \emph{joint I-Q}, indicating a single encoder for both in-phase and quadrature components.} 
    \label{fig:sys}
\end{figure}
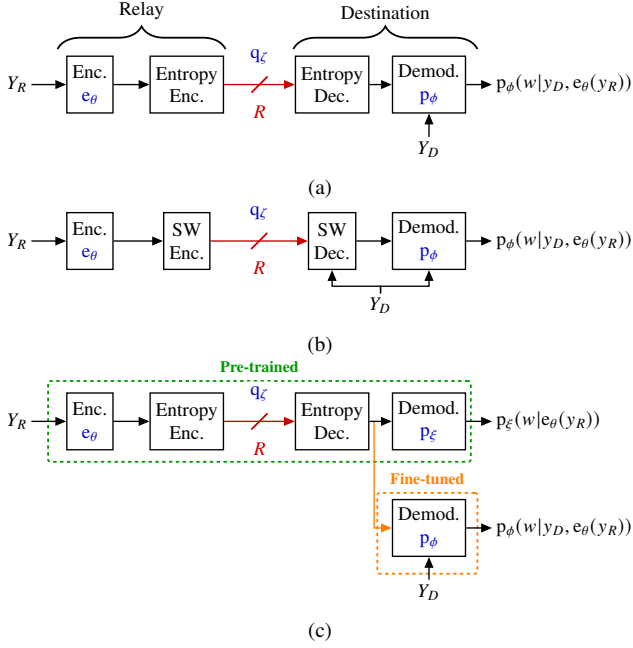 
In this section, by leveraging universal function approximation capability of artificial neural networks (ANNs)~\cite{hornik_et_al, Leshno1993}, we propose  three neural CF schemes to be employed in the PRC shown in Fig.~\ref{fig:sys_model}. We describe these schemes in detail in Sec.~\ref{subsec:neural_cf_architectures} and provide design insights in Sec.~\ref{subsec:motivation_compressor}. Objective function and implementation details are discussed in Sec.~\ref{sec:loss}.
As detailed in in Sec.~\ref{sec:perf}, the modulation scheme remains fixed throughout.
On the other hand, the relay's encoder, employing CF strategy, and the destination's demodulator will be parameterized using ANNs, which will undergo joint optimization in an end-to-end manner.

\subsection{Neural CF Architectures}
\label{subsec:neural_cf_architectures}
Building onto neural distributed compressors proposed in~\cite{Ezgi-ISIT-2023}, we consider learning-based CF schemes that include neural one-shot WZ compressors (with side information $Y_D$ at the destination), paired with either a classic entropy coder (EC) or a SW coder, at the relay. 
We will name these two variants as \emph{marginal} (marg.) and \emph{conditional} (cond.) formulations, respectively.
As a benchmark, we also consider a neural one-shot \emph{point-to-point} (p2p) compressor coupled with a classic EC. 
All of these learned compressors are combined with a neural demodulator available at the destination, which has access to the side information $Y_{D}$.

The overall proposed learned CF relaying architectures are illustrated in Fig.~\ref{fig:sys}.
The encoder's ANN at the relay is denoted by $\Enc(\cdot)$, with $\theta$ representing its parameters; 
the probability distribution of the relay encoder's output (which is then used by the EC or SW coder) is modeled with $\EntMod$, parameterized by $\zeta$;
the demodulator's ANN is $\Demod(w\vert y_D,\Enc(y_R))$, where $\phi$ denotes its parameters.
The mapping defined by the demodulator $\Demod$ represents the posterior probability over the alphabet $\{1,\dots,\vert\mathcal{X}\vert\}$ (soft decision), which serves as an approximation of the true posterior distribution $p(w \vert y_{D}, \Enc(y_{R}))$. 
In the learning process of a point-to-point compressor, as shown in Fig.~\ref{fig:p2p_model}, we initially train a demodulator $\Demodprime(w\vert\Enc(y_R))$ to prevent this neural compressor from utilizing the side information $Y_{D}$ during training. 
The pre-trained point-to-point neural compressor as such (highlighted in green) is then used as input for fine-tuning the demodulator $\Demod(w \vert y_D,\Enc(y_R))$, which incorporates side information (highlighted in orange). 

We set the relay encoder's output as $U \triangleq \Enc(Y_R)$ as in Fig.~\ref{fig:sys_model}. Envisioning a practical scheme,  we have $U$ as discrete.
Specifically, we have that $\Enc(Y_R)\in\{1,...,K\}$, where $K$ is a model parameter. This parameter $K$ is chosen large enough to guarantee sufficient support for the encoder output.
To facilitate the learning process of the encoder, we will use a probabilistic model for $\Enc(Y_R)$ during training. We set the encoder output in a deterministic way, as in~\cite{Ezgi-ISIT-2023}, that is $u= \argmax_{k \in \{1, \dots, K\}}  \Enc (y_{R})$ for a given $Y_{R}=y_R$. Note that the encoder $\Enc$ operates in an unordered \emph{categorical} space, outputting one of the categories of the quantization index $k \in \{1, \dots, K\}$ for each input realization.

Similar to~\cite{Ezgi-ISIT-2023},  without loss of generality, we define the probabilistic models $\Enc(Y_R)$ (during training) and $\EntMod$ as discrete distributions with probabilities as follows:
\begin{equation}
    P_k = \frac{\exp \alpha_k}{\sum_{i=1}^K\exp \alpha_i },
    \label{eq:discrete-P}
\end{equation}
for $k \in \{1, \dots, K\}$. The unnormalized log-probabilities (\emph{logits}) $\alpha_i$ are either directly treated as learnable parameters or computed by ANNs as functions of the conditioning variable. We note that the lossless compression rates induced by the models $\EntMod$ are attainable with high-order classic EC~\cite{universal_modeling} or SW coder~\cite{SW_coding}, operating on discrete values.

For experiments involving complex-valued modulation schemes, the CF architectures depicted in Fig.~\ref{fig:sys} compress the in-phase (i.e. real) and quadrature (i.e. imaginary) components of $Y_R$ \emph{jointly}. We also explore the variants of these architectures illustrated in Fig.~\ref{fig:sys} (not shown), where the in-phase and quadrature components are given as input to two separate encoders, each of which has parameters of its own. The compression rate in this case is computed as the sum of the rates achieved by the in-phase and the quadrature entropy coding schemes (involving either classic EC or SW coders for both). Similar to the architectures outlined in Fig.~\ref{fig:sys}, we still employ a single demodulator $\Demod$ for all these variants, which takes as input the two indices coming from the compressed representations of the in-phase and the quadrature components, along with the side information \(Y_D\).  
We will refer to these architectural configurations as the \emph{split I-Q} variants, whereas we will name the original architectures shown in Fig.~\ref{fig:sys} as the \emph{joint I-Q} versions. 

Although joint compression of in-phase and quadrature components using schemes illustrated in Fig.~\ref{fig:sys} should, in principle, outperform independent processing as in the split I-Q variant, we argue that incorporating domain knowledge into the design, especially when training learning-based schemes, can sometimes facilitate finding the optimal solution for the algorithm. This will be further clarified in Sec.~\ref{subsec:numerical_results}. We refer the readers to Sec.~\ref{sec:loss} for further details on the training procedure.

\subsection{Rationale Behind Our Design Choices}
\label{subsec:motivation_compressor}
While the popular class of neural image compressors (e.g.,~\cite{Balle2017, balle2018variational, BalleJournal}) seems well-suited for distributed compression, and more specifically for the WZ problem, analysis in~\cite{ozyilkan2023neural} reveals that it fails to learn efficient many-to-one mappings exploiting the side information. Consequently, these popular schemes do not recover proper binning schemes, which are known to be optimal in the asymptotic setting~\cite{Wyner_Ziv}, for abstract exemplary sources (such as the quadratic-Gaussian case), severely limiting their compression efficiency.
In~\cite{ozyilkan2024breaking}, it is hypothesized that this limitation stems from the inherent \emph{spectral bias}~\cite{rahaman2019spectral} of the popular class of neural compressors. This spectral bias arises because the encoder outputs operate on the real line. This inherently favors learning smooth functions, consequently hindering these neural compressors from capturing highly discontinuous functions and many-to-one mappings such as binning.

Based on this, our proposed learning-based CF schemes, as in the case of the learned WZ compressors~\cite{Ezgi-ISIT-2023, ozyilkan2023neural}, operate directly within an unordered \emph{categorical} space, similar to traditional vector quantization. Our neural relay compressors are, therefore, in the form of entropy-constrained vector quantizers that can more easily leverage correlated signal available at the destination. Note that this is in contrast to the popular class of neural compressors~\cite{Balle2017, balle2018variational, BalleJournal}, where each of the dimensions at the encoder output is subjected to entropy-constrained scalar quantization in an ordered transform space operating on real line.

The design choices explained in Sec.~\ref{subsec:neural_cf_architectures} maintain the parametric families in their most general form, avoiding any unnecessary imposition of structure. In particular, these would enable the model $\Enc$ to recover, when necessary, quantization schemes featuring discontiguous quantization bins, reminiscent of the \emph{random binning} operation in the achievability of the WZ theorem~\cite{Wyner_Ziv}, which also appears in the CF relaying strategy~\cite{relaycapacity, Simeone_2}.

\subsection{Objective Function}
\label{sec:loss}
In contrast to prior works on neural distributed compression~\cite{Ezgi-ISIT-2023, ozyilkan2023neural}, which focus on minimizing the \emph{distortion} in the reconstruction of the input source in tandem with variable rate entropy coding, our goal in this work is to optimize the operational trade-off between relay-to-destination compression rate and source-to-destination communication rate in the PRC setup, underscoring the task-aware nature of the relay compressor design.

For our objective function, building onto the relay rate in~\eqref{eq:opt2}, we first consider the following upper bound:
\begin{align}
    \mathrm{I}(Y_{R}; \, U \;  \vert \;  Y_{D})  &\leq H(U \; \vert \; Y_{D}), \label{eq:cross_entropy_0}\\
    &\leq  \mathbb{E}\left[ -\log_{2} \EntMod( \Enc(y_{R}))\right] \stackrel{\Delta}{=} \tilde{R}, \label{eq:cross_entropy} 
\end{align}
where $\tilde{R}$ represents an operational upper bound on the relay's \emph{compression} rate, which is limited by $R$.
The inequality in~\eqref{eq:cross_entropy} is due to the fact that the cross-entropy is larger or equal to entropy~\cite[Theorem 5.4.3]{elements_of_information_theory}. 
Here, $\tilde{R}$ encapsulates the compression rate of a relay quantizer having a one-shot encoder coupled with high-order entropy coder over large blocks of the quantized source. 

Similarly, we also establish a lower bound based on the achievable rate  in~\eqref{eq:opt1} as follows:
\begin{align}
    I(X;Y_D, U) 
     & =  H(W)- H(W\;  \vert \; Y_D, U), \label{eq:comm_rate_term_1}  \\
     & \geq   \log(\vert\mathcal{X}\vert)-  \tilde{D}, \label{eq:comm_rate_term_2}
\end{align} 
 where $\tilde{D} \stackrel{\Delta}{=}\mathbb{E} \left[-\log (\Demod(x \vert y_{D}, \Enc(y_{R})))\right]$, 
 and~\eqref{eq:comm_rate_term_2} is a lower bound on the source-to-destination \emph{communication} rate $C$ from~\eqref{eq:opt1}. 
 Here,~\eqref{eq:comm_rate_term_1} follows from $X$ being a one-to-one deterministic function of $W$, and~\eqref{eq:comm_rate_term_2} is again due to cross-entropy being larger or equal to entropy.
 Since we have a fixed modulation scheme and do not perform any probabilistic shaping, we have $H(W)=H(X)=\log(\vert\mathcal{X}\vert)$ in~\eqref{eq:comm_rate_term_2}. 

For a demodulator making hard decisions as: 
\begin{equation}
    \hat{W} = \arg\max_{w\in\{1,\dots,\vert\mathcal{X}\vert\}} \Demod(w\vert y_D,\Enc(y_R)),
\label{eq:hard}
\end{equation}
the corresponding symbol error rate (SER) is defined as:
\begin{equation}
    \text{SER} = P(W \neq \hat{W}).
    \label{eq:SER}
\end{equation}
Since minimizing the cross-entropy $\tilde{D}$ is known to be a surrogate for maximizing the accuracy of classification (that is symbol detection)~\cite{Bishop-book}, minimizing $\tilde{D}$ also operationally corresponds to minimizing SER. 

Building onto the above bounds, the training objective of all the proposed neural CF relaying schemes depicted in Fig.~\ref{fig:sys} can be described by the following loss function: 
\begin{align} 
\begin{split}
    \label{eq:loss_fn} 
    L(\theta,\phi,\zeta) &= \tilde{R} + \lambda \tilde{D}, \\
\end{split}
\end{align}
where $\tilde{R}$ and $\tilde{D}$ are from~\eqref{eq:cross_entropy} and~\eqref{eq:comm_rate_term_2} respectively, and $\lambda > 0$ controls the trade-off. 
The optimized $\Enc$, $\EntMod$ and $\Demod$ models, parameterized by $\theta$, $\zeta$ and $\phi$, yield the ANN-based encoder, EC or SW coder, and demodulator component, respectively. The upper bound in~\eqref{eq:cross_entropy} corresponds to the compression rate achievable by a CF relaying scheme employing a one-shot task-aware encoder $\Enc$ and demodulator $\Demod$, both coupled with an entropy code based on $\EntMod$ (either classic EC or SW coder). This asymptotic compression rate is equivalent to the cross-entropy $\mathbb{E}\left[ -\log_{2} \EntMod( \Enc(y_{R}))\right] $. Similarly, the lower bound in~\eqref{eq:comm_rate_term_2} corresponds to the overall communication rate achieved by a capacity achieving channel code, operating over large blocklengths, used in conjunction with the (soft) demodulator $\Demod$. Therefore, minimizing the loss function in~\eqref{eq:loss_fn} enables the end-to-end optimization of this {\em operational} relaying scheme.

We draw on standard design assumptions in our formulation from both the quantization~\cite{quantization} and neural compression~\cite{BalleJournal} literatures. While our proposed schemes compress each realization of the received relay signal one at a time, employing \emph{one-shot} quantization, we assume that the entropy coding is applied over large blocks of the quantized source elements, i.e., in a \emph{high-order} fashion~\cite{Gish}. This choice is justified by the fact that properly designed entropy codes achieve rates that are only negligibly higher than the entropy values themselves~\cite{universal_coding}.

Consistent with findings in \cite{nested_relay_quantizer, practical_relay_quantizer}, we empirically confirmed that minimizing mean squared error distortion metric at the quantizers may not always maximize the source-to-destination communication rate. The intuition for this is as follows: A distortion-minimizing quantizer aims to preserve \emph{the relay's received signal}, whereas the relay quantizer should instead retain the \emph{source information} as the relay's end goal is to facilitate the communication in the source-to-destination link, highlighting the task-aware nature of the compressor design objective at hand. Grounded in information theoretical principles following~\cite{Simeone_2}, this key insight underlies the objective function (see~\eqref{eq:loss_fn}) for our neural CF relaying schemes.

Adjusting the trade-off parameter $\lambda$ in~\eqref{eq:loss_fn} results different points within the achievable region. The learnable parameters are amenable to joint optimization using stochastic gradient descent (SGD)  since the loss function is differentiable with respect to them. The gradients can be computed using automatic differentiation methods, as implemented in deep learning frameworks such as JAX~\cite{jax}.

As in the popular class of neural compressors~\cite{BalleJournal}, we use SGD to optimize all learnable parameters jointly, which relies on Monte Carlo approximation for the expectations in the loss function.  In SGD, the expectations in the loss functions are replaced by averages over batches of samples $B$, and the order of differentiation and summation is exchanged due to linearity. For a given generic pair of $X=x$ and $Y=y$, let $\ell_{\theta}(x,y)$ denote the sample loss with parameters $\theta$ (represented as one of the sample loss functions inside the brackets in~\eqref{eq:loss_fn}). In this case, Monte Carlo approximation yields:
\begin{align}
    \frac{\partial }{\partial \theta }\mathbb{E} [\;  \ell_{\theta}(x,y)) \; ] \approx \frac{1}{\vert B \vert} \sum_{(x, y)\in B} \frac{\partial \ell_{\theta}(x,y)}{\partial \theta} \; .   \label{eq:sgd}
\end{align} 
This requires that we draw some samples from the model $\Enc$ throughout training. The Gumbel-max trick, initially proposed in~\cite{gumbel_org}, provides a method to draw samples from any discrete distribution. It does so by drawing samples from a distribution of $K$ states (as in~\eqref{eq:discrete-P}) as follows:
\begin{equation} \label{eq:gumbel_max}
    \argmax _{k \in \{1, \dots, K \} }\{\alpha_{k} + G_{k} \},
\end{equation}
where $G_{k}$ are i.i.d. samples from a standard Gumbel distribution. 

Recognizing that the derivative of the $\argmax$ operator in~\eqref{eq:gumbel_max} is zero everywhere except at the boundaries of state changes, we opt for a \emph{continuous relaxation} of this operator during training to carry out SGD. Such a relaxation is provided by the Concrete distribution, introduced in~\cite{concrete}. Rather than obtaining discrete (hard) samples, this method produces soft samples, forming a vector of length $K$ where the mass is distributed across multiple states instead of being concentrated in one. The index $k \in \{1, \dots, K\}$ of such a soft sample is determined using a \emph{softmax} function:
\begin{equation}
    U_k = \frac{\exp((\alpha_{k}+G_k)\; / \; t)}{\sum_{i=1}^K\exp((\alpha_{i}+G_i)\; /\; t)} \; ,\label{eq:softmax}
\end{equation}
where $t$ is a temperature parameter that controls the amount of relaxation. As $t \rightarrow 0^{+}$, the soft samples converge to their hard counterparts, indicating that the Concrete distribution converges to a discrete one. Throughout training, we also choose the Concrete distribution for the models $\EntMod$ to match the distribution of samples from $\Enc$.

During evaluation, we transition from Concrete distributions back to their discrete counterparts. As explained in Sec.~\ref{subsec:neural_cf_architectures},
we also use a deterministic encoding function equivalent to the mode of $\Enc$, instead of sampling from it, by setting encoder output as $u= \argmax_{k \in \{1, \dots, K\}}  \Enc (y_{R})$.

Note that in spite of considering specific modulation schemes in training, we do not assume \emph{a priori} knowledge of the modulation scheme by the relay in our neural CF schemes. 
The parameters $\{\theta,\phi,\zeta\}$ are learned solely in a data-driven fashion from samples, through the proposed loss function in~\eqref{eq:loss_fn}. Similarly, the relay also has no prior information on the channel gains $h_R$ and $h_D$ (see~\eqref{eq:prc}).
Further improvement in the performance may be obtained by also learning an optimized probabilistic shaping ($p(x)$ in optimization~\eqref{eq:opt1}-\eqref{eq:opt2}) and a geometric shaping (constellation $\mathcal{X}$) of the modulation~\cite{Stark-2019}. 

\section{Results and Discussion} 
\label{sec:discussion}
While our framework can be adapted to different modulation schemes and PRC setups, we adopt the following system configuration to showcase numerical results.
As stated in Sec.~\ref{sec:system_model}, we assume equally likely symbols, i.e., $p(x)=1/\vert\mathcal{X}\vert$.
The average power constraint on the transmitted signal is $\E[\vert X \vert^2]= P$.
For real-valued channels, we consider BPSK, 4-PAM, and 8-PAM modulations, having constellations $\Xcal=\{\pm A\}$, $\Xcal=\{\pm A, \pm 3A\}$, and $\Xcal=\{\pm A, \pm 3A, \pm 5A, \pm 7A\}$, respectively, where $A$ is chosen to satisfy the power constraint $P$.  
For complex-valued channels, we consider 4-QAM and 16-QAM modulations with power constraint $P$.
We recall that the SNR at the destination and at the relay is defined as $\gamma_D = \vert h_D\vert^2 P$ and $\gamma_R = \vert h_R\vert^2 P$, respectively.

For the parametrization of $\Enc$ and $\Demod$, we use ANNs of three dense layers, with 100 units each, except the last one, and leaky rectified linear unit as the activation function. In our experiments, we observed that increasing the size of the networks or employing different activation functions did not lead to improved results. The demodulator $\Demod$ receives a concatenated vector comprising both its inputs, $\Enc(Y_R)$ and $Y_{D}$.

We perform our experiments using the JAX framework~\cite{jax} and employ Adam~\cite{adam}, a widely used variant of SGD. We use a learning rate of $10^{-4}$, which we chose by monitoring the convergence of the loss function in the high-rate regime, reducing it by a constant factor of 10 each time the loss visibly plateaued. We found that convergence of the high-rate models takes longer than low-rate models, so we simply carried over our schedule to the lower-rate cases. All neural CF schemes are trained for 500 epochs with randomly initialized network weights. We use a batch size of $B=1024$ (as in~\eqref{eq:sgd}) and set the model parameter $K=32$. The output dimension of $\Demod$ is set to be $\vert\mathcal{X}\vert$, since this probabilistic model represents the posterior over the transmitted constellation. 

We evaluate our learned CF relaying schemes in terms of the trade-off between the relay rate $R$ (using the proxy $\tilde{R}$ in~\eqref{eq:cross_entropy}), and two metrics:
(i) the communication rate $\mathrm{I}(X;Y_D, U)$, for which we use the lower bound (hence, a pessimistic estimate) in~\eqref{eq:comm_rate_term_2}, and 
(ii) the $\text{SER} = P(W \neq \hat{W})$ (see~\eqref{eq:SER}). All empirical estimates of compression rates, communication rates, and bit error rates are obtained by averaging over at least \(10^{6}\)  source realizations.

The rest of this section is organized as follows.
Baseline references for $R=0$ and $R\to\infty$ are presented in~\ref{sec:results-baselines}.
The performance of various learned CF relay schemes is analyzed in~\ref{subsec:numerical_results}, while an interpretation of the corresponding relay's encoder and destination's demodulator is provided in~\ref{sec:results-viz}.
Finally, results for robustness against different SNRs are shown in~\ref{sec:robustness}. 

\subsection{Baselines}
\label{sec:results-baselines}
The regimes where $R=0$ and $R\to\infty$ are referred to as \emph{without relay} and \emph{perfect relay} scenario, respectively. 
When $R=0$, the destination has only access to $Y_D$, having an effective SNR of $\gamma_D$.
In the perfect relay regime ($R\to\infty$), however, the destination has full access to both $Y_D$ and $Y_R$, and it optimally combines them.
This effectively results in an increased SNR of $\gamma_D+\gamma_R$ compared to the scenario without a relay. 
In these two regimes, mutual information and SER can be numerically computed for the considered modulations as a function of $(\gamma_D,\gamma_R)$~\cite{proakis2008digital}. 

When $0<R<\infty$, we consider $C_\text{CF}$ from~\eqref{eq:C_CF-Simeone} (or its complex channel equivalent) as a benchmark for the achievable communication rate of our learned CF schemes with discrete modulations.
Increasing the modulation order, $\vert\mathcal{X}\vert$, gives more degrees of freedom for the end-to-end learned communication system to approach the rate of a PRC that assumes Gaussian inputs, as represented by $C_\text{CF}$ in~\eqref{eq:C_CF-Simeone}.

\subsection{Performance of the Learned CF Relaying Schemes}
\label{subsec:numerical_results}
For the first set of results, we assume that the SNR is the same for both the destination and the relay, i.e., $\gamma_D = \gamma_R$.

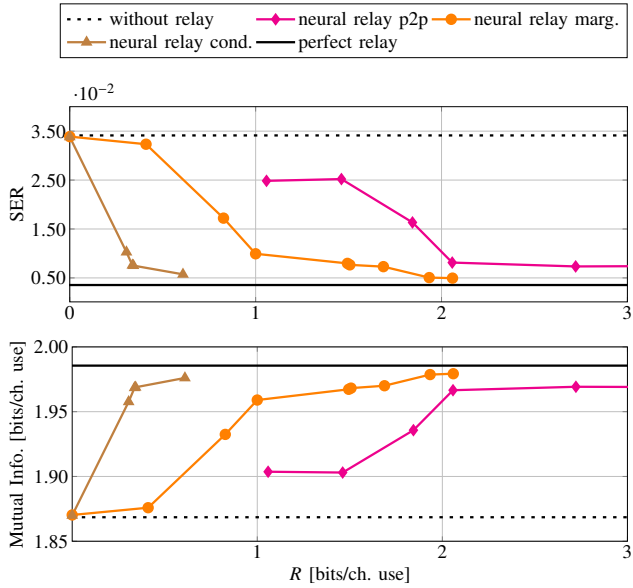
\begin{figure}
    \centering
    \pgfplotsset{
    compat=newest,
    /pgfplots/legend image code/.code={%
        \draw[mark repeat=2,mark phase=2,#1] 
            plot coordinates {
                (0cm,0cm) 
                (0.6cm,0cm)
                (1.2cm,0cm)%
            };
    },
}

\begin{tikzpicture}
\begin{axis}[
	scale only axis,
	font=\LARGE,
	width=\textwidth,
	height=0.35*\textwidth,
	xmajorgrids=true,
	ymajorgrids=true,
	xmin=0,
	xmax=3,
        xtick={0,1, ..., 3},
	ymin=0.0001,
	ymax=0.04,
        ytick={0.005,0.015,...,0.04},
        y tick label style={
        /pgf/number format/.cd,
            fixed,
            fixed zerofill,
            precision=2,
        /tikz/.cd
    },
	ylabel={SER},
	legend columns=3,
	legend cell align={left},
	legend style={row sep=3pt, at={(1,1.25)}, anchor = south east, inner sep =5pt},
	]

        \addplot+[line width=2pt, mark=none, black, loosely dashed] table[x=X, y=Y, col sep=tab] 
    {fig-1571030310/data/4PAM_noRelay.txt};

    \addplot+[line width=2pt, magenta, solid, mark=diamond*, mark options={scale=2,fill=magenta,solid}] table[x=X, y=Y, col sep=space] 
    {fig-1571030310/data/4PAM_p2p.txt};

    \addplot+[line width=2pt, orange, solid, mark=*, mark options={scale=2,fill=orange,solid}] table[x=X, y=Y, col sep=space] 
    {fig-1571030310/data/4PAM_marg.txt};
 
	\addplot+[line width=2pt, brown, solid, mark=triangle*, mark options={scale=2,fill=brown,solid}] table[x=X, y=Y, col sep=space] 
    {fig-1571030310/data/4PAM_cond.txt};

    \addplot+[line width=2pt, mark=none, black, solid] table[x=X, y=Y, col sep=tab] 
    {fig-1571030310/data/4PAM_perfRelay.txt};

    \legend{
        without relay,
        neural relay p2p,
        neural relay marg.,
        neural relay cond.,
        perfect relay
    }
\end{axis}
\end{tikzpicture}\\
    \vspace{0.5em}
    \pgfplotsset{
    compat=newest,
    /pgfplots/legend image code/.code={%
        \draw[mark repeat=2,mark phase=2,#1] 
            plot coordinates {
                (0cm,0cm) 
                (0.6cm,0cm)
                (1.2cm,0cm)%
            };
    },
}

\begin{tikzpicture}
\begin{axis}[
	scale only axis,
	font=\LARGE,
	width=\textwidth,
	height=0.35*\textwidth,
	xmajorgrids=true,
	ymajorgrids=true,
	xmin=0,
	xmax=3,
        xtick={1,2,3},
	ymin=1.85,
	ymax=2.0,
        y tick label style={
        /pgf/number format/.cd,
            fixed,
            fixed zerofill,
            precision=2,
        /tikz/.cd
    },
    ytick={1.8, 1.85, 1.9, 1.95, 2.0},
	xlabel={$R$ [bits/ch. use]},
        ylabel={Mutual Info.  [bits/ch. use]},
	legend columns=1,
	legend cell align={left},
	legend style={row sep=3pt, at={(1,0)}, anchor = south east},
	]
	
	\addplot+[line width=2pt, mark=none, black, loosely dashed] table[x=X, y=Z, col sep=tab] 
    {fig-1571030310/data/4PAM_noRelay.txt};

    \addplot+[line width=2pt, magenta, solid, mark=diamond*, mark options={scale=2,fill=magenta,solid}] table[x=X, y=Z, col sep=space] 
    {fig-1571030310/data/4PAM_p2p.txt};

    \addplot+[line width=2pt, orange, solid, mark=*, mark options={scale=2,fill=orange,solid}] table[x=X, y=Z, col sep=space] 
    {fig-1571030310/data/4PAM_marg.txt};
 
	\addplot+[line width=2pt, brown, solid, mark=triangle*, mark options={scale=2,fill=brown,solid}] table[x=X, y=Z, col sep=space] 
    {fig-1571030310/data/4PAM_cond.txt};

    \addplot+[line width=2pt, mark=none, black, solid] table[x=X, y=Z, col sep=tab] 
    {fig-1571030310/data/4PAM_perfRelay.txt};

\end{axis}

\end{tikzpicture}
    \caption{Symbol error rate (SER) and mutual information as a function of the relay-to-destination rate $R$,  for the 4-PAM modulation with $\gamma_D = \gamma_R = 13$ dB.
    The colored lines represent the performance of three neural CF relay architectures (Fig.~\ref{fig:sys}), where each marker corresponds to a unique model trained for a particular value of $\lambda$ in~\eqref{eq:loss_fn}.
    The horizontal black lines provide baseline results without relaying ($R=0$) and with perfect relaying ($R\to\infty$).
    }
    \label{fig:rate_vs_overall_performance}
\end{figure}
    
Fig.~\ref{fig:rate_vs_overall_performance} shows the SER and mutual information for the 4-PAM modulation 
when $\gamma_D = \gamma_R = 13$ dB.
In this case, $Y_R$ and $Y_D$, are highly correlated. We observe that the three models exhibit different trade-offs. 
Recall that, in the point-to-point variant depicted in Fig.~\ref{fig:p2p_model}, $\Enc$ is not able to use $Y_{D}$ as side information in compression.
As seen in Fig.~\ref{fig:rate_vs_overall_performance}, the conditional model yields the best performance as the side information is also exploited within the SW coder. 
The marginal model surpasses the point-to-point model mainly due to exploiting the side information during compression (see Sect.~\ref{sec:results-viz} for a more detailed discussion), yielding rate reduction.

\begin{figure}
    \centering
    \pgfplotsset{
    compat=newest,
    /pgfplots/legend image code/.code={%
        \draw[mark repeat=2,mark phase=2,#1] 
            plot coordinates {
                (0cm,0cm) 
                (0.6cm,0cm)
                (1.2cm,0cm)%
            };
    },
}

\begin{tikzpicture}
\begin{axis}[
	scale only axis,
	font=\LARGE,
	width=\textwidth,
	height=0.5*\textwidth,
	xmajorgrids=true,
	ymajorgrids=true,
	xmin=0,
	xmax=10,
        xtick={0,1, ..., 10},
	ymin=0.2,
	ymax=0.43,
        ytick={0.2,0.25,0.3,...,0.5},
        y tick label style={
        /pgf/number format/.cd,
            fixed,
            fixed zerofill,
            precision=2,
        /tikz/.cd
    },
	ylabel={SER},
	legend columns=2,
	legend cell align={left},
	legend style={row sep=3pt, at={(1,1.1)}, anchor = south east, inner sep =5pt},
	]

        \addplot+[line width=2pt, mark=none, black, loosely dashed] table[x=X, y=Y, col sep=tab] 
    {
X	Y
0	4.19E-01
20	4.19E-01
    };

    \addplot+[line width=2pt, mark=none, black, solid] table[x=X, y=Y, col sep=tab] 
    {
X	Y
0	2.21E-01
20	2.21E-01
    };

    \addplot+[line width=2pt, magenta, solid, mark=diamond*, mark options={scale=2,fill=magenta,solid}] table[x=X, y=Y, col sep=space] 
    {
X	Y
0	0.4193601608
3.029412031	0.3065567017
3.809128761	0.2907218933
4.915579796	0.2768144608
6.525149345	0.2714414597
9.391876221	0.2632875443
    };

    \addplot+[line width=2pt, magenta, dotted, mark=diamond*, mark options={scale=2,fill=magenta,solid}] table[x=X, y=Y, col sep=space] 
    {
X	Y
0	0.4193601608
2.054406881	0.3707017899
3.548602343	0.3425207138
4.861817837	0.3134393692
5.711174011	0.2792196274
7.093050003	0.2406377792
7.915064335	0.2375869751
8.596953392	0.2361841202
9.733801842	0.2302923203
10.59508133	0.2292575836
11.00697517	0.2278652191
    };

    \addplot+[line width=2pt, orange, solid, mark=*, mark options={scale=2,fill=orange,solid}] table[x=X, y=Y, col sep=space] 
    {
X	Y
1.79E-05	0.4190292358
8.67E-01	0.3871278763
2.21E+00	0.315703392
3.44E+00	0.2768344879
4.18E+00	0.2608604431
4.656479359	0.2555456161
    };

    \addplot+[line width=2pt, orange, dotted, mark=*, mark options={scale=2,fill=orange,solid}] table[x=X, y=Y, col sep=space] 
    {
X	Y
0.000022014	0.4190568924
1.938979149	0.3500165939
3.568590164	0.2982091904
5.168862343	0.2549695969
5.973480701	0.2430591583
6.982209683	0.2369003296
7.596447945	0.2351284027
8.028189659	0.2332363129
8.519053459	0.2305040359
9.1388731	0.2298002243
    };

	\addplot+[line width=2pt, brown, solid, mark=triangle*, mark options={scale=2,fill=brown,solid}] table[x=X, y=Y, col sep=space] 
    {
X	Y
0.00E+00	0.4192028
1.44E+00	0.31955528
1.95E+00	0.28938007	
2.4356198	0.27215004
2.904616	0.2583847
    };

    \addplot+[line width=2pt, brown, dotted, mark=triangle*, mark options={scale=2,fill=brown,solid}] table[x=X, y=Y, col sep=space] 
    {
X	Y
0.0042943237	0.4190006256
1.025192976	0.359992981
2.133359194	0.2864742279
2.529007912	0.2723875046
3.020807028	0.2630710602
3.668181896	0.2501058578
4.297627449	0.2413635254
4.860966682	0.2362670898
5.605315685	0.2326602936
6.39889288	0.2300720215
    };

    \legend{
        without relay,         
        perfect relay,
        neural relay p2p (joint I-Q),
        neural relay p2p (split I-Q),
        neural relay marg. (joint I-Q),
        neural relay marg. (split I-Q),
        neural relay cond. (joint I-Q),
        neural relay cond. (split I-Q)
    }
\end{axis}
\end{tikzpicture}\\
    \vspace{0.5em}
    \pgfplotsset{
    compat=newest,
    /pgfplots/legend image code/.code={%
        \draw[mark repeat=2,mark phase=2,#1] 
            plot coordinates {
                (0cm,0cm) 
                (0.6cm,0cm)
                (1.2cm,0cm)%
            };
    },
}

\begin{tikzpicture}
\begin{axis}[
	scale only axis,
	font=\LARGE,
	width=\textwidth,
	height=0.5*\textwidth,
	xmajorgrids=true,
	ymajorgrids=true,
	xmin=0,
	xmax=10,
        xtick={0,1,2,3,...,10},
	ymin=2.25,
	ymax=3.25,
        y tick label style={
        /pgf/number format/.cd,
            fixed,
            fixed zerofill,
            precision=2,
        /tikz/.cd
    },
    ytick={2.25,2.5, ..., 3.5},
	xlabel={$R$ [bits/ch. use]},
        ylabel={Mutual Info. [bits/ch. use]},
	legend columns=1,
	legend cell align={left},
	legend style={row sep=3pt, at={(1,0)}, anchor = south east},
	]
	
        \addplot+[line width=2pt, mark=none, black, loosely dashed] table[x=X, y=Y, col sep=tab] 
    {
X	Y
0	2.44E+00
20	2.44E+00
    };

    \addplot+[line width=2pt, mark=none, black, solid] table[x=X, y=Y, col sep=tab] 
    {
X	Y
0	3.17E+00
20	3.17E+00
    };

    \addplot+[line width=2pt, magenta, solid, mark=diamond*, mark options={scale=2,fill=magenta,solid}] table[x=X, y=Y, col sep=space] 
    {
X	Y
0	2.438678503
3.029412031	2.862217903
3.809128761	2.912111044
4.915579796	2.956810236
6.525149345	2.975611925
9.391876221	3.00353241
    };

    \addplot+[line width=2pt, magenta, dotted, mark=diamond*, mark options={scale=2,fill=magenta,solid}] table[x=X, y=Y, col sep=space] 
    {
X	Y
0	2.438678503
2.054406881	2.624479294
3.548602343	2.728054047
4.861817837	2.834197044
5.711174011	2.951363802
7.093050003	3.090952635
7.915064335	3.101993084
8.596953392	3.107736588
9.733801842	3.129227638
10.59508133	3.133259535
11.00697517	3.137861252
    };

    \addplot+[line width=2pt, orange, solid, mark=*, mark options={scale=2,fill=orange,solid}] table[x=X, y=Y, col sep=space] 
    {
X	Y
1.79E-05	2.439818382	
8.67E-01	2.553782463
2.21E+00	2.806143999
3.44E+00	2.944237709
4.18E+00	3.006083489
4.656479359	3.034600496
    };

    \addplot+[line width=2pt, orange, dotted, mark=*, mark options={scale=2,fill=orange,solid}] table[x=X, y=Y, col sep=space] 
    {
X	Y
0.000022014	2.438361645
1.938979149	2.687814236
3.568590164	2.867187262
5.168862343	3.034579516
5.973480701	3.077613592
6.982209683	3.10653615
7.596447945	3.112594366
8.028189659	3.121572971
8.519053459	3.130638123
9.1388731	3.131943941
    };

	\addplot+[line width=2pt, brown, solid, mark=triangle*, mark options={scale=2,fill=brown,solid}] table[x=X, y=Y, col sep=space] 
    {
X	Y
0.00E+00	2.4381247
1.44E+00	2.807747
1.95E+00	2.9134254
2.4356198	2.9756732
2.904616	3.0205266
    };

    \addplot+[line width=2pt, brown, dotted, mark=triangle*, mark options={scale=2,fill=brown,solid}] table[x=X, y=Y, col sep=space] 
    {
X	Y
0.0042943237	2.438062906
1.025192976	2.663721085
2.133359194	2.927137375
2.529007912	2.974622726
3.020807028	3.009537935
3.668181896	3.050875187
4.297627449	3.089359522
4.860966682	3.106963158
5.605315685	3.120660782
6.39889288	3.132234573
    };

\end{axis}

\end{tikzpicture}
    \caption{
       Symbol error rate (SER) and mutual information as a function of the relay-to-destination rate $R$,  for the 16-QAM modulation with $\gamma_D = \gamma_R = 7$ dB.
    The colored lines illustrate the performance of three neural CF relay architectures depicted in Fig.~\ref{fig:sys}, accompanied by their respective \emph{split I-Q} variants (as introduced in Sec.~\ref{subsec:neural_cf_architectures}). In the figure, each marker corresponds to a unique model trained for a specific value of $\lambda$ in~\eqref{eq:loss_fn}.
    The horizontal black lines indicate baseline results without relaying ($R=0$) and with perfect relaying ($R\to\infty$).
    }
    \label{fig:rate_vs_overall_performance-QAM}
\end{figure}
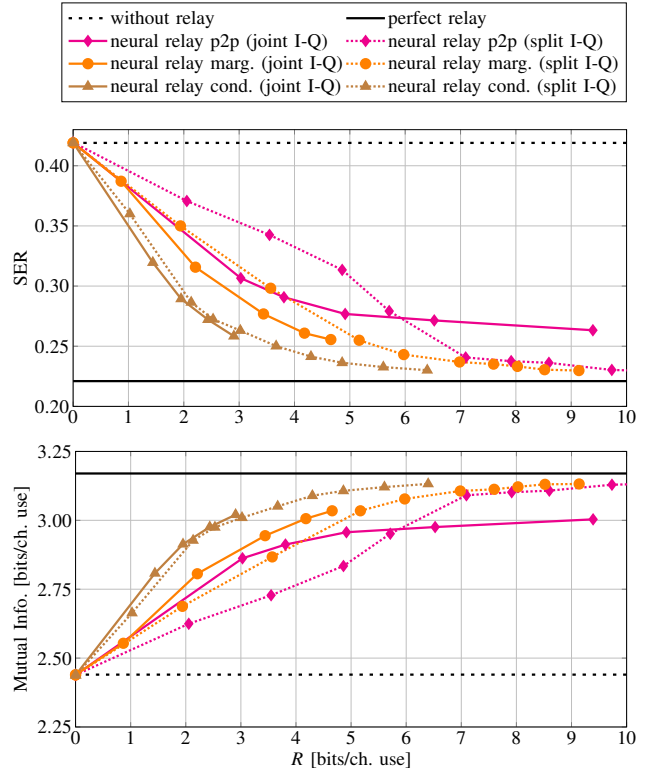

Similarly, Fig.~\ref{fig:rate_vs_overall_performance-QAM} contains the SER and the mutual information for 16-QAM modulation when $\gamma_D = \gamma_R = 7$ dB.
In this case, we also show results where two separate encoders compress the in-phase and quadrature part of $Y_R$ independently -- these models are annotated as \emph{split I-Q} variants, as introduced in Sec.~\ref{subsec:neural_cf_architectures}. We observe that at lower rates, the architectures depicted in Fig.~\ref{fig:sys}, which correspond to {\em joint I-Q} compression, perform best across three different schemes (conditional, marginal and point-to-point). In these models, both the in-phase and quadrature components are fed into a single encoder $\Enc$, enabling joint compression of real and imaginary parts of the complex-valued input signal. This allows these compressors to learn more flexible quantization boundaries (not depicted), making them more efficient in the low-rate regime. However, at higher rates, we observe that the split I-Q variants outperform their joint counterparts. 
Since one would expect ``grid-like'' quantization boundaries for QAM modulations, imposing separate processing on real and imaginary parts in the split I-Q models, effectively leverages this domain knowledge, enabling them to approach capacity at high rates. 
In contrast, all of the joint I-Q architectures for conditional, marginal and point-to-point variants saturate around a capacity value of 3. These results suggest that as the modulation order and relay rate increase, incorporating domain knowledge into the compressor design could be beneficial. Imposing such well-informed design structures in this case further enhances the efficiency of learned CF relaying schemes, particularly at high rates, where training neural compressors becomes relatively more challenging compared to the low rate regime.

\begin{figure}
    \centering
    \pgfplotsset{
    compat=newest,
    /pgfplots/legend image code/.code={%
        \draw[mark repeat=2,mark phase=2,#1] 
            plot coordinates {
                (0cm,0cm) 
                (0.6cm,0cm)
                (1.2cm,0cm)%
            };
    },
}

\begin{tikzpicture}
\begin{axis}[
	scale only axis,
	font=\LARGE,
	width=\textwidth,
	height=0.5*\textwidth,
	xmajorgrids=true,
	ymajorgrids=true,
	xmin=0,
	xmax=4,
        xtick={0,1,...,10},
	ymin=0.7,
	ymax=1.2,
        y tick label style={
        /pgf/number format/.cd,
            fixed,
            fixed zerofill,
            precision=2,
        /tikz/.cd
    },
    ytick={0.0, 0.1, ..., 2.0},
	xlabel={$R$ [bits/ch. use]},
        ylabel={Mutual Info. [bits/ch. use]},
	legend columns=1,
	legend cell align={left},
	legend style={row sep=3pt, at={(1,0)}, anchor = south east},
	]
	


    \addplot+[line width=2pt, mark=none, black, solid] table[x=X, y=Y, col sep=tab] {
X	Y
0	0.7924812504
0.1	0.8308342149
0.2	0.8659678294
0.3	0.8980118147
0.4	0.9271159153
0.5	0.9534452978
0.6	0.9771758998
0.7	0.9984899674
0.8	1.017571968
0.9	1.034605009
1	1.049767837
1.1	1.063232442
1.2	1.07516225
1.3	1.085710859
1.4	1.095021256
1.5	1.103225439
1.6	1.110444366
1.7	1.116788171
1.8	1.122356567
1.9	1.127239385
2	1.131517203
2.1	1.135262028
2.2	1.138537992
2.3	1.141402048
2.4	1.143904646
2.5	1.146090376
2.6	1.147998572
2.7	1.149663872
2.8	1.151116738
2.9	1.152383917
3	1.153488877
3.1	1.15445218
3.2	1.155291832
3.3	1.156023589
3.4	1.156661225
3.5	1.157216779
3.6	1.157700765
3.7	1.158122365
3.8	1.158489589
3.9	1.158809428
4	1.15908798
};

\addplot+[line width=2pt, \myRed, solid, mark=triangle*, mark options={scale=2,fill=\myRed,solid}] table[x=X, y=Y, col sep=tab] {
X	Y
7.51E-04	0.77594197
5.29E-02	0.7816848
6.01E-02	0.7828052
9.98E-01	0.95298123
1.56E+00	1.023695
2.30E+00	1.0762223
2.95E+00	1.0962241
3.709006	1.1124394
};

\addplot+[line width=2pt, \myBlue, solid, mark=*, mark options={scale=2,fill=\myBlue,solid}] table[x=X, y=Y, col sep=tab] {
X	Y
4.22E-04	0.7716543
1.5697752	1.0185666
2.3182433	1.0694047
2.9231832	1.087146
3.289085	1.0924349
3.806848	1.099398
4.055839	1.1003636
    };

    \addplot+[line width=2pt, \myGreen, solid, mark=square*, mark options={scale=2,fill=\myGreen,solid}] table[x=X, y=Y, col sep=tab] {
X	Y
0	0.7218482
0.9994287	0.87024045
1.3683649	0.8897412
1.8483359	0.9014961
2.4205663	0.9079051
};

    \addplot+[line width=2pt, \myGreen, mark=none, dotted] table[x=X, y=Y, col sep=tab] {
X	Y
1	0.9128222858
5	0.9128222858
};

    \addplot+[line width=2pt, \myBlue, mark=none, dotted] table[x=X, y=Y, col sep=tab] {
X	Y
2	1.10302412
5	1.10302412
};

    \addplot+[line width=2pt, \myRed, mark=none, dotted] table[x=X, y=Y, col sep=tab] {
X	Y
2	1.117926487
5	1.117926487
};

    \addplot+[line width=2pt, black, mark=none, dotted] table[x=X, y=Y, col sep=tab] {
X	Y
2	1.160964047
5	1.160964047
};

\legend{
    $C_\text{{CF}}$ from~\eqref{eq:C_CF-Simeone},
    8-PAM neural relay marg.,
    4-PAM neural relay marg.,
    BPSK neural relay marg.,
}

\end{axis}

\end{tikzpicture}
    \caption{
        Mutual information for the marginal model (Fig.~\ref{fig:marg_model}) in case of BPSK, 4-PAM and 8-PAM modulations with $\gamma_D = \gamma_R = 3$ dB.
        The solid line represents $C_\text{CF}$ in~\eqref{eq:C_CF-Simeone}~\cite{Simeone_2}, obtained for Gaussian inputs.
        The dotted lines represent the perfect relay ($R\to\infty$) bounds for the respective curves, 
        similar to Figs.~\ref{fig:rate_vs_overall_performance} and~\ref{fig:rate_vs_overall_performance-QAM}.
    }
    \label{fig:rate_vs_mutual_info}
\end{figure}
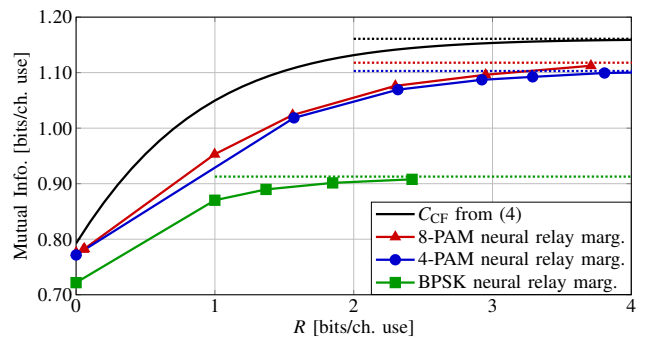

Fig.~\ref{fig:rate_vs_mutual_info} compares $C_\text{CF}$ from~\eqref{eq:C_CF-Simeone} with the mutual information obtained with the marginal formulation (Fig.~\ref{fig:marg_model}) for the BPSK, 4-PAM and 8-PAM modulations.
Here, the SNR for all the considered schemes is $\gamma_D = \gamma_R = 3$ dB, suggesting a lower correlation between $Y_R$ and $Y_D$ compared to the one illustrated in Fig.~\ref{fig:rate_vs_overall_performance}. As expected, increasing the modulation order narrows the gap to the bound in~\eqref{eq:C_CF-Simeone}. Notably, the marginal variant meets the performance of the corresponding perfect relay ($R\to\infty$) baseline at higher rates.

\begin{figure}
    \centering
    \pgfplotsset{
    compat=newest,
    /pgfplots/legend image code/.code={%
        \draw[mark repeat=2,mark phase=2,#1] 
            plot coordinates {
                (0cm,0cm) 
                (0.6cm,0cm)
                (1.2cm,0cm)%
            };
    },
}

\begin{tikzpicture}
\begin{axis}[
	scale only axis,
	font=\LARGE,
	width=\textwidth,
	height=0.5*\textwidth,
	xmajorgrids=true,
	ymajorgrids=true,
	xmin=0,
	xmax=5,
        xtick={0,1,...,10},
	ymin=0.85,
	ymax=2.75,
        y tick label style={
        /pgf/number format/.cd,
            fixed,
            fixed zerofill,
            precision=2,
        /tikz/.cd
    },
    ytick={0.0, 0.25, ..., 4.0},
	xlabel={$R$ [bits/ch. use]},
        ylabel={Mutual Info. [bits/ch. use]},
	legend columns=1,
	legend cell align={left},
	legend style={row sep=3pt, at={(1,0.1)}, anchor = south east},
	]
	
    \addplot+[line width=2pt, mark=none, black, solid] table[x=X, y=Y, col sep=tab] {
X	Y
0	2.196158711
0.2	2.286829871
0.4	2.364039769
0.6	2.428605493
0.8	2.481700841
1	2.524714855
1.2	2.559112149
1.4	2.586318625
1.6	2.607643156
1.8	2.624234731
2	2.637067998
2.2	2.646948181
2.4	2.65452715
2.6	2.660324448
2.8	2.664749224
3	2.668120749
3.2	2.670686426
3.4	2.672636942
3.6	2.674118678
3.8	2.675243654
4	2.676097397
4.2	2.676745087
4.4	2.677236332
4.6	2.677608849
4.8	2.677891292
5	2.678105418
5.2	2.678267737
};

\addplot+[line width=2pt, \myRed, solid, mark=triangle*, mark options={scale=2,fill=\myRed,solid}] table[x=X, y=Y, col sep=tab] {
X	Y
2.06E-05	2.054766
1.79E-05	2.05739
0.0014145428	2.0560455
1.0001438	2.184465
2.0867722	2.3788073
2.5109885	2.4110076
3.1460054	2.4389045
3.4739327	2.443806
3.8441398	2.4555008
4.087475	2.4654574
4.4514513	2.4705641
    };

    \addplot+[line width=2pt, \myBlue, solid, mark=*, mark options={scale=2,fill=\myBlue,solid}] table[x=X, y=Y, col sep=tab] {
X	Y
0	1.8702732
0.9998943	1.9589856
1.6873574	1.9700237
2.0590758	1.9792962
3.402115107	1.983902812
};

    \addplot+[line width=2pt, \myGreen, solid, mark=square*, mark options={scale=2,fill=\myGreen,solid}] table[x=X, y=Y, col sep=tab] {
X	Y
0.0005991016	0.9999738336
1.146314383	0.9999964237
};

    \addplot+[line width=2pt, \myBlue, mark=none, dotted] table[x=X, y=Y, col sep=tab] {
X	Y
0.5	1.985544814
5	1.985544814
};

    \addplot+[line width=2pt, \myRed, mark=none, dotted] table[x=X, y=Y, col sep=tab] {
X	Y
2	2.48E+00
5	2.48E+00
};

    \addplot+[line width=2pt, \myGreen, mark=none, dotted] table[x=X, y=Y, col sep=tab] {
X	Y
0	0.9999999994
5	0.9999999994
};

    \addplot+[line width=2pt, black, mark=none, dotted] table[x=X, y=Y, col sep=tab] {
X	Y
2	2.678776002
5	2.678776002
};

\legend{
    $C_\text{CF}$ from~\eqref{eq:C_CF-Simeone},
    8-PAM neural relay marg.,
    4-PAM neural relay marg.,
    BPSK neural relay marg.,
}

\end{axis}

\end{tikzpicture}
    \caption{
        Mutual information  for the marginal model (Fig.~\ref{fig:marg_model}) in case of BPSK, 4-PAM and 8-PAM modulations with $\gamma_D = \gamma_R = 13$ dB.
        The solid line represents $C_\text{CF}$ in~\eqref{eq:C_CF-Simeone}~\cite{Simeone_2}, obtained for Gaussian inputs.
        The dotted lines represent the perfect relay ($R\to\infty$) bounds for the respective curves,
        similar to Figs.~\ref{fig:rate_vs_overall_performance}, and~\ref{fig:rate_vs_overall_performance-QAM}.
    }
    \label{fig:rate_vs_mutual_info-13dB}
\end{figure}

Similarly, Fig.~\ref{fig:rate_vs_mutual_info-13dB} compares $C_\text{CF}$ from~\eqref{eq:C_CF-Simeone} for the same modulation schemes considered in Fig.~\ref{fig:rate_vs_mutual_info}, but now at higher SNR $\gamma_R=\gamma_D = 13$ dB, suggesting a stronger correlation between $Y_{R}$ and $Y_{D}$. We note that at higher SNR, as theory suggests, the rate allowed by higher-order modulation is greater and the performance gap between different modulation schemes is larger. At high rates, for each of the modulation schemes considered, our neural CF schemes once again 
match the communication rate bound for perfect relay ($R\to\infty$), mirroring the trend observed in Fig.~\ref{fig:rate_vs_mutual_info}.

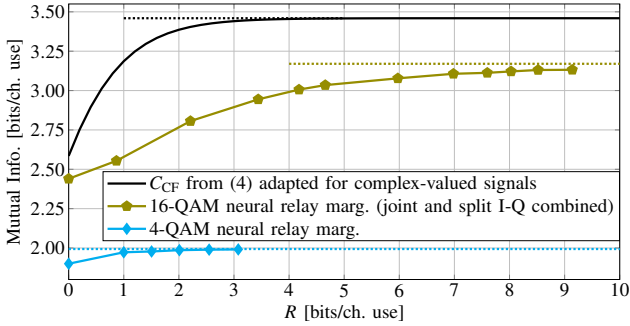
\begin{figure}
    \centering
    \pgfplotsset{
    compat=newest,
    /pgfplots/legend image code/.code={%
        \draw[mark repeat=2,mark phase=2,#1] 
            plot coordinates {
                (0cm,0cm) 
                (0.6cm,0cm)
                (1.2cm,0cm)%
            };
    },
}

\begin{tikzpicture}
\begin{axis}[
	scale only axis,
	font=\LARGE,
	width=\textwidth,
	height=0.5*\textwidth,
	xmajorgrids=true,
	ymajorgrids=true,
	xmin=0,
	xmax=10,
        xtick={0,1,...,10},
	ymin=1.8,
	ymax=3.55,
        y tick label style={
        /pgf/number format/.cd,
            fixed,
            fixed zerofill,
            precision=2,
        /tikz/.cd
    },
    ytick={0.0, 0.25, ..., 4},
	xlabel={$R$ [bits/ch. use]},
        ylabel={Mutual Info. [bits/ch. use]},
	legend columns=1,
	legend cell align={left},
	legend style={row sep=3pt, at={(1,0.14)}, anchor = south east},
	]

    \addplot+[line width=2pt, mark=none, black, solid] table[x=X, y=Y, col sep=tab] {
X	Y
0	2.584962501
0.2	2.753189683
0.4	2.895176076
0.6	3.012932506
0.8	3.109050379
1	3.186413124
1.2	3.247936976
1.4	3.296375967
1.6	3.334200625
1.8	3.363541796
2	3.386182637
2.2	3.403581019
2.4	3.416907717
2.6	3.427090105
2.8	3.434855068
3	3.440767774
3.2	3.445264965
3.4	3.44868256
3.6	3.451278017
3.8	3.45324812
4	3.454742973
4.2	3.455876892
4.4	3.456736836
4.6	3.457388894
4.8	3.457883258
5	3.458258028
5.2	3.458542116
5.4	3.458757451
5.6	3.458920666
5.8	3.459044373
6	3.459138131
6.2	3.459209191
6.4	3.459263047
6.6	3.459303864
6.8	3.459334797
7	3.459358241
7.2	3.459376009
7.4	3.459389474
7.6	3.459399679
7.8	3.459407413
8	3.459413274
8.2	3.459417716
8.4	3.459421082
8.6	3.459423634
8.8	3.459425567
9	3.459427032
9.2	3.459428143
9.4	3.459428985
9.6	3.459429622
9.8	3.459430106
10	3.459430472
};

\addplot+[line width=2pt, olive, solid, mark=pentagon*, mark options={scale=2,fill=olive,solid}] table[x=X, y=Y, col sep=tab] {
X	Y
1.79E-05	2.439818382	
8.67E-01	2.553782463
2.21E+00	2.806143999
3.44E+00	2.944237709
4.18E+00	3.006083489
4.656479359	3.034600496
5.973480701	3.077613592
6.982209683	3.10653615
7.596447945	3.112594366
8.028189659	3.121572971
8.519053459	3.130638123
9.1388731	3.131943941
};

\addplot+[line width=2pt, cyan, solid, mark=diamond*, mark options={scale=2,fill=cyan,solid}] table[x=X, y=Y, col sep=tab] {
X	Y
2.20E-05	1.900237
0.9999887	1.9728982
1.504518	1.978056
2.009083	1.9863076
2.549069	1.9896339
3.0776596	1.9910369
    };

    \addplot+[line width=2pt, olive, mark=none, dotted] table[x=X, y=Y, col sep=tab] {
X	Y
4	3.17E+00
10	3.17E+00
};

    \addplot+[line width=2pt, cyan, mark=none, dotted] table[x=X, y=Y, col sep=tab] {
X	Y
0	1.99359473
10	1.99359473
};

    \addplot+[line width=2pt, black, mark=none, dotted] table[x=X, y=Y, col sep=tab] {
X	Y
1	3.459431619
5	3.459431619
};

\legend{
    $C_\text{CF}$ from~\eqref{eq:C_CF-Simeone} adapted for complex-valued signals,
    16-QAM neural relay marg.
    (joint and split I-Q combined),
    4-QAM neural relay marg.,
}

\end{axis}

\end{tikzpicture}
    \caption{
       Mutual information for the marginal model (Fig.~\ref{fig:marg_model}) in case of 4-QAM and 16-QAM modulations with $\gamma_D = \gamma_R = 7$ dB.
        The solid line represents $C_\text{CF}$ in~\eqref{eq:C_CF-Simeone}~\cite{Simeone_2}, obtained for Gaussian inputs.
        The dotted lines represent the perfect relay ($R\to\infty$) bounds for the respective curves,
        similar to Figs.~\ref{fig:rate_vs_overall_performance} and~\ref{fig:rate_vs_overall_performance-QAM}. In the case of the 16-QAM modulation shown here, we select the best performing model among two variants of the marginal neural relay quantizers: the joint I-Q scheme depicted in Fig.~\ref{fig:marg_model} and the split I-Q architecture introduced in Sec.~\ref{subsec:neural_cf_architectures}, both of which are plotted in orange in Fig.~\ref{fig:rate_vs_overall_performance-QAM}.
    }
    \label{fig:rate_vs_mutual_info-QAM}
\end{figure}

Considering next a complex-valued communication scenario, we illustrate the results obtained with 4-QAM and 16-QAM modulations in Fig.~\ref{fig:rate_vs_mutual_info-QAM}, where we set $\gamma_R=\gamma_D = 7$ dB. For this case, we use an adapted version of $C_\text{CF}$ from~\eqref{eq:C_CF-Simeone} that considers instead a complex-valued PRC. For the 16-QAM results depicted, we select the best performing variant for the marginal model (either the joint I-Q scheme shown in Fig.~\ref{fig:marg_model} or the split I-Q version, both of which are introduced in Sec.~\ref{subsec:neural_cf_architectures}). Consistent with the trends observed in Figs.~\ref{fig:rate_vs_mutual_info} and ~\ref{fig:rate_vs_mutual_info-13dB}, our neural CF scheme again meet the respective perfect relay baselines ($R\to\infty$). 
These empirical results further confirm that our learning-based relay compression schemes can be easily adapted to any chosen fixed modulation, scoring higher communication throughput as the order of modulation increases.

\subsection{Interpretability of the Learned CF Relaying Schemes}
\label{sec:results-viz}
The maximum a posteriori (MAP) estimator for $W$ in the PRC of Fig.~\ref{fig:sys_model} is as follows:
\begin{align}
    \hat{w} &= \arg\max_w p(w \vert y_D,u),\\
    &= \arg\max_w p(y_D\vert w) \; p(u\vert w) \; p(w),
    \label{eq:map}
\end{align}
where we have used the independence of $y_D$ and $u$ given $w$.
For equally likely symbols, $p(w)$ is a constant and therefore, can be removed from~\eqref{eq:map}.
Note that the term $p(u\vert w)$ represents the likelihood of $w$ based on the relay's quantized observation $u$, and it \emph{updates} the destination's likelihood $p(y_D\vert w)$ in~\eqref{eq:map}.
For reference, without the relay, the optimal decision thresholds on $Y_D$ for the maximum likelihood estimator for a PAM (QAM) modulation under Gaussian noise would be the intersection between adjacent likelihoods, that is the middle point (line) between adjacent symbols~\cite{proakis2008digital}.

We recall that for our learned CF schemes introduced in Sec.~\ref{sec:neural_schemes}, $u=\Enc(y_R)$, and the posterior estimated by the neural demodulator is $\Demod(w\vert y_D, \Enc(y_R))$. 
In the remainder of this section, we provide results that help to visualize and interpret the quantization boundaries recovered by the neural encoder $\Enc$ and learned MAP~\eqref{eq:hard} decision thresholds adopted by the demodulator.
First, we show that the marginal CF variant (Fig.~\ref{fig:marg_model}) groups the quantized indices at the relay, by assigning the same quantization index to discontiguous intervals in the source space. This empirical evidence suggests that the scheme effectively uses the side information $Y_D$ during compression.
Next, we show how the relay's likelihood $p(\Enc(y_R) \vert w)$ operationally shifts the decision thresholds.

\begin{figure}
    \centering
    \includegraphics[width=0.95\columnwidth]{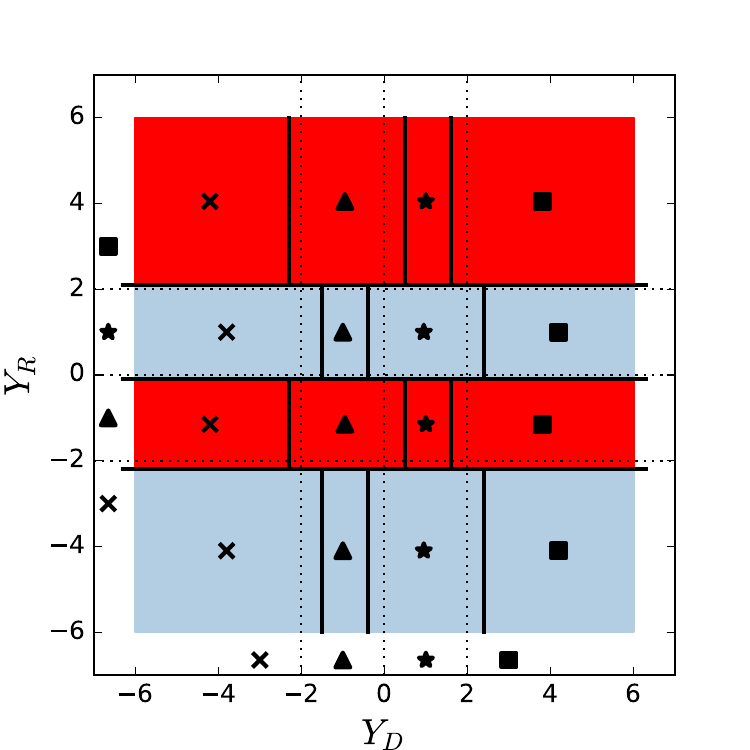}
    \caption{
        Visualization (best viewed in color) of the learned CF strategy (marginal scheme in Fig.~\ref{fig:marg_model}) and demodulation decisions for the 4-PAM modulation with $\gamma = 13$ and relay rate $R\approx 1$.
        The horizontal lines denote the quantization boundaries on $Y_R$, and the colors designate the transmitted index $\Enc(Y_R)$.
        The vertical lines denote the hard decision boundaries for the demodulator, and the markers represent the decisions.
        The transmitted symbols (denoted by cross, triangle, star, square) are also reported near the axis for reference. 
    }
    \label{fig:binning_vis}
\end{figure}

Fig.~\ref{fig:binning_vis} illustrates the marginal CF scheme and the demodulation's hard decision regions (see~\eqref{eq:hard}) for 4-PAM with $\gamma_D = \gamma_D = 13$ dB and relay rate of $R\approx1$.
The vertical axis and horizontal axis show $Y_R$ and $Y_D$, respectively.
The colors represent the transmitted indices $\Enc(Y_R)$ by the relay, and the horizontal lines are the corresponding quantization boundaries.
Note that this neural CF architecture exhibits binning (grouping) since non-adjacent intervals are assigned to the same index (same color). 
It is worth noting that this recovered grouping behavior is similar to the random binning operation in the achievability proof of the WZ theorem~\cite{Wyner_Ziv} and also in the achievability of CF~\cite{relaycapacity}. This emergence of learned one-shot binning behavior also explains the further reduction in relay rate compared to the point-to-point model, as illustrated in the experimental results shown in Figs.~\ref{fig:rate_vs_overall_performance} and~\ref{fig:rate_vs_overall_performance-QAM}. Unlike the marginal scheme, the point-to-point model (Fig.~\ref{fig:p2p_model}), however, lacks access to the side information signal $Y_D$, which is available at the decoder, during compression. Therefore, this latter model cannot learn a binning behavior in the relay compressor (not depicted). In contrast, the conditional variant (Fig.~\ref{fig:cond_model})
leverages the side information not only during compression but also within the entropy coding stage. This enables the conditional scheme to execute binning over long sequences i.e., in a multi-shot fashion. Note that such a high-order binning scheme, facilitated by the SW coder, is inherently more efficient than the one-shot binning achievable by an encoder at the relay. As the model $\Enc$ compresses each source realization one at a time, it can only bin the quantized indices at the relay in a one-shot fashion. 

The vertical lines in Fig.~\ref{fig:binning_vis} denote the hard decision boundaries, where the markers denote the decisions $\hat{W}$. 
We observe that the decision boundaries are shifted with respect to the midpoints between transmitted symbols (optimal boundaries without relaying).
This highlights the interpretability of our neural CF relaying scheme. 
For example, when \emph{cross} or \emph{star} are transmitted, the index \emph{blue} will be the (most likely) relayed index. 
In this case, the decision regions for \emph{cross} and \emph{star} at the destination are larger than the other symbols.

\begin{figure}
    \centering
    \begin{subfigure}[b]{\columnwidth}
        \centering
        \includegraphics[width=0.8\columnwidth]{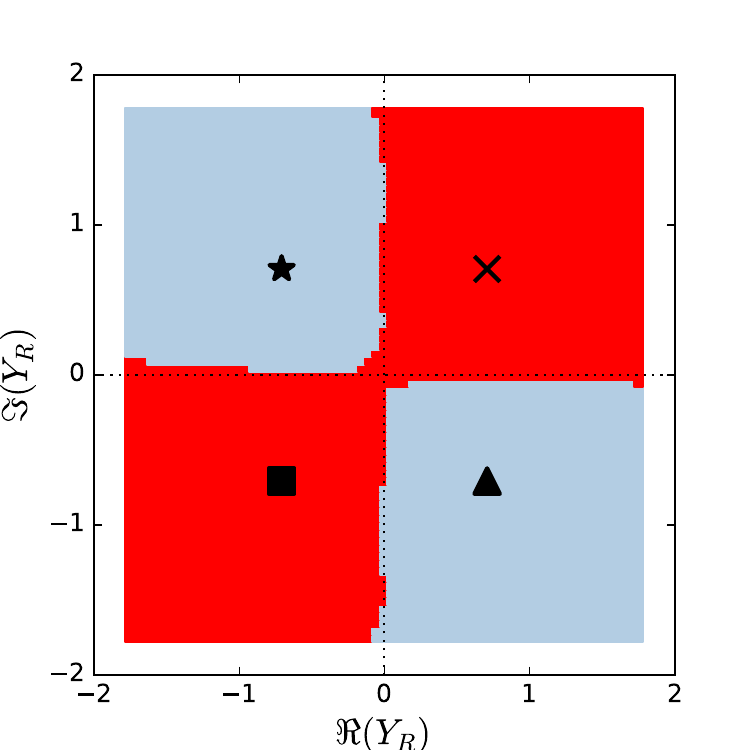}
        \caption{}
        \label{fig:QAM-binning} 
    \end{subfigure}
    \\ 
    \vspace{1em}
    \begin{subfigure}[b]{0.49\columnwidth}
        \centering
        \includegraphics[width=\columnwidth]{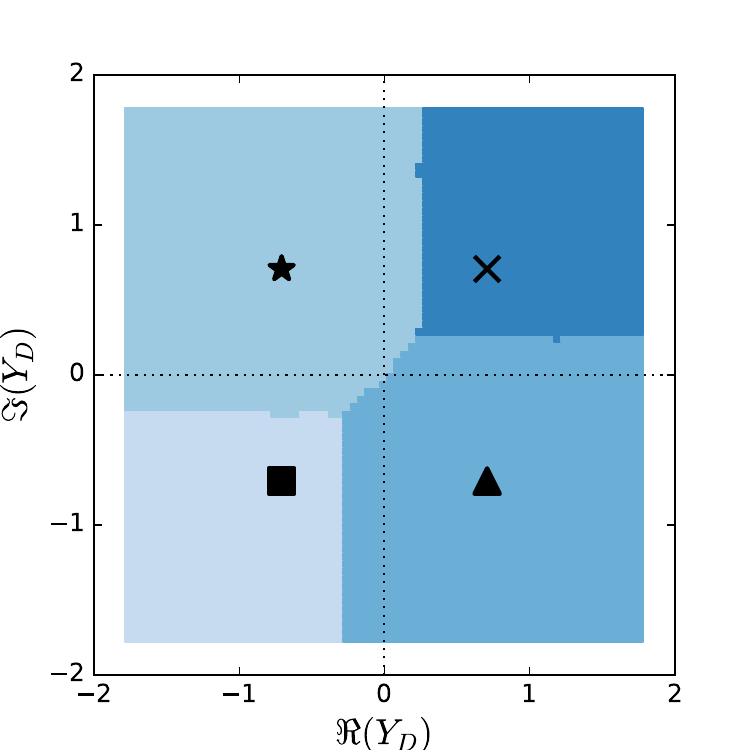}
        \caption{}
        \label{fig:QAM-decision-1} 
    \end{subfigure}
    \begin{subfigure}[b]{0.49\columnwidth}
        \centering
        \includegraphics[width=\columnwidth]{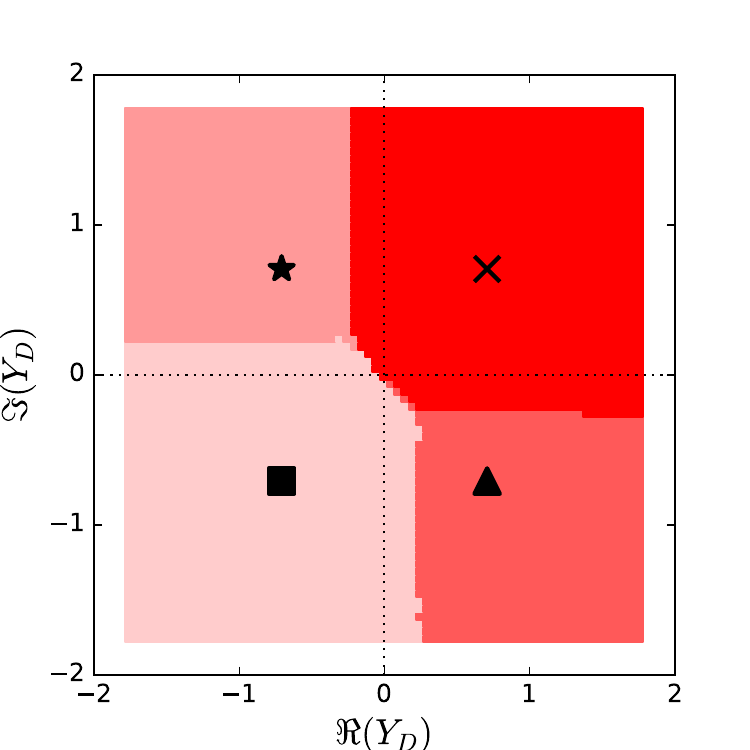}
        \caption{}
        \label{fig:QAM-decision-2} 
    \end{subfigure}
    \caption{
        Visualization (best viewed in color) of the learned CF strategy (marginal scheme in Fig.~\ref{fig:marg_model}) and demodulation decisions for the 4-QAM modulation with $\gamma = 7$ dB and relay rate $R\approx 1$.
        Figure (a) shows the quantization boundaries on $Y_R$ (on the complex plane), and the colors designate the transmitted index $\Enc(Y_R)$.
        Figures (b) and (c) show the hard decision boundaries for the demodulator as a function of $Y_D$ (on the complex plane), where different colors represent the different decisions.
        Figure (b) represents the decisions when $\Enc(Y_R)$ corresponds to the \emph{blue} index from Figure (a);
        Figure (c) represents the decisions when $\Enc(Y_R)$ corresponds to the \emph{red} index from Figure (a).
        The transmitted symbols (denoted by cross, triangle, star, square) are also reported for reference. 
    }
    \label{fig:binning_vis-QAM}
\end{figure} 

Fig.~\ref{fig:binning_vis-QAM} shows the learned marginal CF strategy for the complex-valued 4-QAM modulation when $\gamma_D = \gamma_R =7$ dB and relay rate of $R\approx1$.
The vertical and horizontal axis of each subfigure represent real and imaginary parts of $Y_R$ and $Y_D$.
Fig.~\ref{fig:QAM-binning} reports the output of the relay's encoder, where the color represents $\Enc(Y_R)$.
One can note that the regions surrounding the farthest symbols are paired with the same encoding (color) $\Enc(Y_R)$.
Similar to Fig.~\ref{fig:binning_vis}, one can argue that this is yet another instance of binning in the relay compressor.
Fig.~\ref{fig:QAM-decision-1} shows the hard decision boundaries on $Y_D$ when $\Enc(Y_R)$ corresponds to the \emph{blue} index. Meanwhile, Fig.~\ref{fig:QAM-decision-2} shows the hard decision boundaries on $Y_D$ when $\Enc(Y_R)$ corresponds to the \emph{red} index.
Again, the decision boundaries are shifted to favor the symbols that were most likely to be received at the relay. 

In practice, Figs.~\ref{fig:binning_vis} and~\ref{fig:binning_vis-QAM} can be used as  look-up tables for direct deployment of the resulting CF relaying strategies, including both the relay's encoder and the destination's demodulator.
Although ANN-based architectures (Fig.~\ref{fig:sys}) were used to minimize the loss function in~\eqref{eq:loss_fn}, the actual CF scheme and the hard demodulator implementation at test time rely only on the learned quantization boundaries and threshold values shown in Figs.~\ref{fig:binning_vis} and~\ref{fig:binning_vis-QAM}. 

\subsection{Robustness to Signal-to-Noise Ratio (SNR) Variations}
\label{sec:robustness}
In Sections~\ref{sec:results-baselines},~\ref{subsec:numerical_results}, and~\ref{sec:results-viz}, we evaluated the performance at the same SNRs used for training. In this section, we analyze robustness with respect to the training SNR.
We consider 4-PAM modulation for the source $X$, and a range of test SNRs $\gamma_D,\gamma_R\in\{0,1,\dots,6\}$ dB. 
We consider models that satisfy the relay rate constraint of $R \lessapprox 1$.
Note that for this SNR range, it is known that the 4-PAM capacity is superior to the BPSK one, and almost equivalent to those achieved by higher order PAM modulations~\cite{proakis2008digital}. This is also evident in Fig.~\ref{fig:rate_vs_mutual_info} at $\gamma_D=\gamma_R=3$ dB.
Adapting the modulation order to the SNR is a key component of modern communication systems relying on link adaption~\cite{adaptiveMQAM-1997}, and as such, we assume 4-PAM modulation is only used in the above SNR range.

We study the following test scenarios:
\begin{enumerate}
    \item Same SNR at both the relay and the destination, i.e., $\gamma_D=\gamma_R=\gamma\in\{0,1,\dots,6\}$ dB;     
    \item Relay SNR fixed at $\gamma_R=3$ dB, and variable destination SNR $\gamma_D\in\{0,1,\dots,6\}$ dB;
    \item Destination SNR fixed at $\gamma_D=3$ dB, variable relay SNR $\gamma_R\in\{0,1,\dots,6\}$ dB.
\end{enumerate}
For all of the tests above, we consider baseline models that are trained at a single SNR $\gamma_D=\gamma_R=\gamma$, where $\gamma\in\{0,1,\dots,6\}$ dB.
In the remainder of this subsection, we analyze the performance of the baselines on the abovementioned scenarios, and propose alternative learning strategies based on robust training.

\subsubsection{Testing at the same SNR at both the relay and the destination}

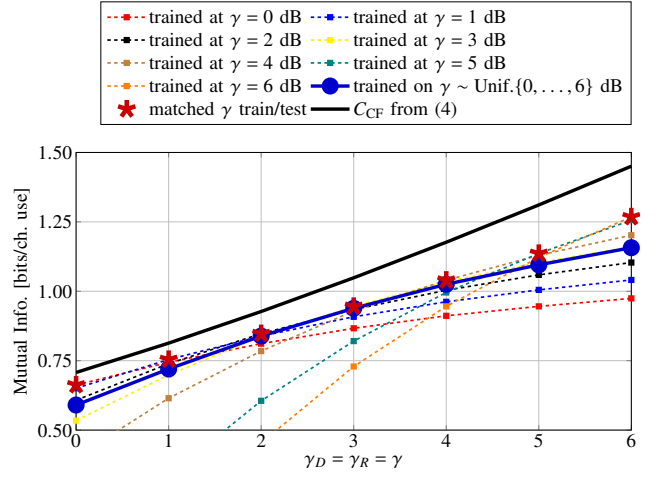
\begin{figure}[t]
    \centering
    \pgfplotsset{
    compat=newest,
    /pgfplots/legend image code/.code={%
        \draw[mark repeat=2,mark phase=2,#1] 
            plot coordinates {
                (0cm,0cm) 
                (0.6cm,0cm)
                (1.2cm,0cm)%
            };
    },
}

\begin{tikzpicture}
\begin{axis}[
	scale only axis,
	font=\LARGE,
	width=\textwidth,
	height=0.5*\textwidth,
	xmajorgrids=true,
	ymajorgrids=true,
	xmin=0,
	xmax=6,
        xtick={0,1,2,3,...,10},
	ymin=0.5,
	ymax=1.5,
        y tick label style={
        /pgf/number format/.cd,
            fixed,
            fixed zerofill,
            precision=2,
        /tikz/.cd
    },
    ytick={0,0.25,0.5,...,10},
	xlabel={$\gamma_D=\gamma_R=\gamma$},
        ylabel={Mutual Info. [bits/ch. use]},
	legend columns=2,
	legend cell align={left},
	legend style={row sep=3pt, at={(1,1.1)}, anchor = south east, inner sep =5pt},
    cycle list name=color list,
	]

    \addplot+[line width=1.5pt, dashed, mark=square*, mark options={scale=1,fill,solid}] table[x=X, y=Y, col sep=tab] 
    {
X	Y
0	0.6625125408
1	0.7443355322
2	0.8112875223
3	0.8666170239
4	0.9116575122
5	0.9456444979
6	0.9743909836
    };

    \addplot+[line width=1.5pt, dashed, mark=square*, mark options={scale=1,fill,solid}] table[x=X, y=Y, col sep=tab] 
    {
X	Y
0	0.6506360769
1	0.75418818
2	0.8384269476
3	0.9080915451
4	0.9621316195
5	1.005031824
6	1.040815115
    };

    \addplot+[line width=1.5pt, dashed, mark=square*, mark options={scale=1,fill,solid}] table[x=X, y=Y, col sep=tab] 
    {
X	Y
0	0.6075966954
1	0.7391675115
2	0.8473067284
3	0.9341685176
4	1.003754377
5	1.059129953
6	1.103525043
    };

    \addplot+[line width=1.5pt, dashed, mark=square*, mark options={scale=1,fill,solid}] table[x=X, y=Y, col sep=tab] 
    {
X	Y
0	0.5341864824
1	0.6986709833
2	0.837204814
3	0.9453676939
4	1.034461975
5	1.103317499
6	1.160499811
    };

    \addplot+[line width=1.5pt, dashed, mark=square*, mark options={scale=1,fill,solid}] table[x=X, y=Y, col sep=tab] 
    {
X	Y
0	0.4027647972
1	0.6147700548
2	0.7846458554
3	0.9288857579
4	1.039811969
5	1.129727483
6	1.201949954
    };

    \addplot+[line width=1.5pt, dashed, mark=square*, mark options={scale=1,fill,solid}] table[x=X, y=Y, col sep=tab] 
    {
X	Y
0	0.01440770738
1	0.3352162242
2	0.6054645777
3	0.8206598163
4	0.9956968427
5	1.136159301
6	1.256172299
    };

    \addplot+[line width=1.5pt, dashed, mark=square*, mark options={scale=1,fill,solid}] table[x=X, y=Y, col sep=tab] 
    {
X	Y
0	-0.2874189317
1	0.1173753366
2	0.4538640976
3	0.7291103601
4	0.9457577467
5	1.122087717
6	1.267825484
    };

    \addplot+[line width=3pt, mark=*, \myBlue, solid, mark options={scale=3,fill=\myBlue,solid}] table[x=X, y=Y, col sep=tab] 
    {
X	Y
0	0.5903180838
1	0.7202681303
2	0.8387856483
3	0.9387693405
4	1.025303125
5	1.095203996
6	1.156979799
    };

    \addplot+[line width=4pt, \myRed, solid, mark=star, only marks, mark options={scale=4,fill=\myRed,solid}] table[x=X, y=Y, col sep=tab] 
    {
X	Y
0	0.6625125408
1	0.75418818
2	0.8473067284
3	0.9453676939
4	1.039811969
5	1.136159301
6	1.267825484
    };

    \addplot+[line width=3pt, mark=none, black, solid] table[x=X, y=Y, col sep=tab] 
    {
X	Y
0	0.7075202199
1	0.8132574908
2	0.9271144724
3	1.048482397
4	1.176667905
5	1.310929733
6	1.450521151
    };

    \legend{
        trained at $\gamma=0$ dB,
        trained at $\gamma=1$ dB,
        trained at $\gamma=2$ dB,
        trained at $\gamma=3$ dB,
        trained at $\gamma=4$ dB,
        trained at $\gamma=5$ dB,
        trained at $\gamma=6$ dB,
        trained on {$\gamma\sim \text{Unif.}\{0,\dots,6\}$} dB,
        matched $\gamma$ train/test,
        $C_\text{CF}$ from~\eqref{eq:C_CF-Simeone}
    }
    
\end{axis}

\end{tikzpicture}
    \caption{
        Robustness analysis when the destination and the relay have the same test signal-to-noise ratio (SNR) $\gamma_D=\gamma_R = \gamma$.
        The rate constraint is $R\approx 1$ for all points.
        The lines represent the mutual information obtained with the learned CF strategy (marginal scheme in Fig.~\ref{fig:marg_model}), as a function of the testing SNR $\gamma$.
        The dotted lines represent models trained for a single value of $\gamma$.
        The solid blue line represents the model trained for \emph{robustness} over the SNR range of interest, i.e., the training SNR is $\gamma \sim \text{Unif.}\{0,1,\dots,6\}$ dB.
        The red stars represent the points where testing and training SNR match. 
    }
    \label{fig:robustness-sameSNR}
\end{figure}

Fig.~\ref{fig:robustness-sameSNR} shows the mutual information when the same SNR is experienced at both the destination and the relay, i.e., $\gamma_D=\gamma_R=\gamma$.
The rate constraint is satisfied for all the models $R\approx 1$ (not shown here).
We also include a \emph{robust} model trained on a range of SNRs $\gamma_D=\gamma_R=\gamma \sim\text{Unif.}\{0,1,\dots,6\}$ dB.
Note that the baseline models trained at a single SNR perform well for adjacent SNRs too.
The robust model trained on the range $\gamma \in \{0,1,\dots,6\}$ dB exhibits a good compromise, offering performance similar to the model trained for the SNR in the middle of the range, and minimal performance degradation in the lower and higher end of the SNR range.
Another observation is that models trained for lower SNRs exhibit less degradation at higher SNRs compared to the opposite case; in fact, the models trained at high SNRs fail at lower SNRs. 

\subsubsection{Testing on a range of SNRs at the destination, fixing the SNR at the relay}

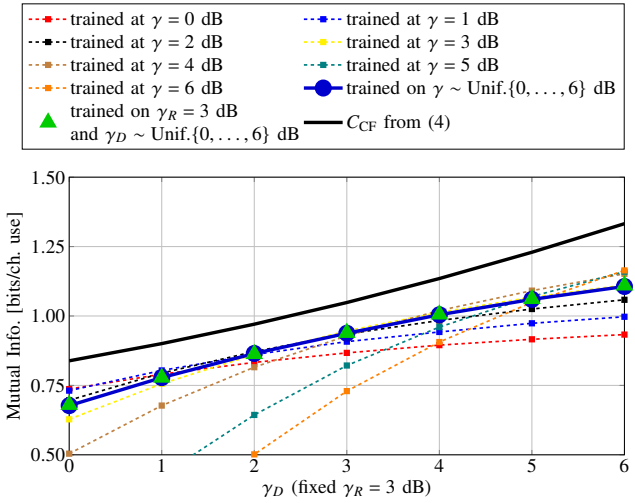
\begin{figure}[t]
    \centering
    \pgfplotsset{
    compat=newest,
    /pgfplots/legend image code/.code={%
        \draw[mark repeat=2,mark phase=2,#1] 
            plot coordinates {
                (0cm,0cm) 
                (0.6cm,0cm)
                (1.2cm,0cm)%
            };
    },
}

\begin{tikzpicture}
\begin{axis}[
	scale only axis,
	font=\LARGE,
	width=\textwidth,
	height=0.5*\textwidth,
	xmajorgrids=true,
	ymajorgrids=true,
	xmin=0,
	xmax=6,
        xtick={0,1,2,3,...,10},
	ymin=0.5,
	ymax=1.5,
        y tick label style={
        /pgf/number format/.cd,
            fixed,
            fixed zerofill,
            precision=2,
        /tikz/.cd
    },
    ytick={0,0.25, 0.5,...,10},
	xlabel={$\gamma_D$ (fixed $\gamma_R=3$ dB)},
        ylabel={Mutual Info. [bits/ch. use]},
	legend columns=2,
	legend cell align={left},
	legend style={row sep=3pt, at={(1,1.1)}, anchor = south east, inner sep =5pt, cells={align=left}},
    cycle list name=color list,
	]

    \addplot+[line width=1.5pt, dashed, mark=square*, mark options={scale=1,fill,solid}] table[x=X, y=Y, col sep=tab] 
    {
X	Y
0	0.7367767096
1	0.7903381586
2	0.8327690363
3	0.8670896292
4	0.8950093985
5	0.9159961939
6	0.9330114126
    };

    \addplot+[line width=1.5pt, dashed, mark=square*, mark options={scale=1,fill,solid}] table[x=X, y=Y, col sep=tab] 
    {
X	Y
0	0.7310996056
1	0.8038908839
2	0.8609876633
3	0.9076017141
4	0.9424204826
5	0.9736577272
6	0.9971920848
    };

    \addplot+[line width=1.5pt, dashed, mark=square*, mark options={scale=1,fill,solid}] table[x=X, y=Y, col sep=tab] 
    {
X	Y
0	0.6964531541
1	0.7941493988
2	0.8718267679
3	0.935038209
4	0.9838039279
5	1.025107861
6	1.058316588
    };

    \addplot+[line width=1.5pt, dashed, mark=square*, mark options={scale=1,fill,solid}] table[x=X, y=Y, col sep=tab] 
    {
X	Y
0	0.6279004216
1	0.7559693456
2	0.861076951
3	0.9463024139
4	1.01299417
5	1.067884088
6	1.111441374
    };

    \addplot+[line width=1.5pt, dashed, mark=square*, mark options={scale=1,fill,solid}] table[x=X, y=Y, col sep=tab] 
    {
X	Y
0	0.5041217208
1	0.6773155332
2	0.8155174255
3	0.9283472896
4	1.01960516
5	1.09130156
6	1.152038336
    };

    \addplot+[line width=1.5pt, dashed, mark=square*, mark options={scale=1,fill,solid}] table[x=X, y=Y, col sep=tab] 
    {
X	Y
0	0.152670905
1	0.425624758
2	0.6436287761
3	0.8216177821
4	0.9600373507
5	1.070155382
6	1.15963912
    };

    \addplot+[line width=1.5pt, dashed, mark=square*, mark options={scale=1,fill,solid}] table[x=X, y=Y, col sep=tab] 
    {
X	Y
0	-0.1356428564
1	0.2099790573
2	0.5015203953
3	0.7294174433
4	0.906162262
5	1.048951268
6	1.164181948
    };

    \addplot+[line width=3pt, mark=*, \myBlue, solid, mark options={scale=3,fill=\myBlue,solid}] table[x=X, y=Y, col sep=tab] 
    {
X	Y
0	0.6771844625
1	0.7780331969
2	0.8649219275
3	0.9377601743
4	1.004425287
5	1.059790969
6	1.106271267
    };

    \addplot+[line width=4pt, green!80!black, solid, mark=triangle*, only marks, mark options={scale=2.5,fill=green!80!black,solid}] table[x=X, y=Y, col sep=tab] 
    {
X	Y
0	0.680110395
1	0.7782539129
2	0.8632723689
3	0.9394245148
4	1.005741715
5	1.061438918
6	1.110186338
    };

    \addplot+[line width=3pt, mark=none, black, solid] table[x=X, y=Y, col sep=tab] 
    {
X	Y
0	0.8385259931
1	0.9005191636
2	0.9704197753
3	1.048481819
4	1.134851103
5	1.229530636
6	1.332390895
    };

    \legend{
        trained at $\gamma=0$ dB,
        trained at $\gamma=1$ dB,
        trained at $\gamma=2$ dB,
        trained at $\gamma=3$ dB,
        trained at $\gamma=4$ dB,
        trained at $\gamma=5$ dB,
        trained at $\gamma=6$ dB,
        trained on {$\gamma \sim \text{Unif.}\{0,\dots,6\}$} dB,
        trained on {$\gamma_R=3$ dB}\\ {and $\gamma_D \sim \text{Unif.}\{0,\dots,6\}$} dB,
        $C_\text{CF}$ from~\eqref{eq:C_CF-Simeone}
    }
    
\end{axis}

\end{tikzpicture}
    \caption{
        Robustness analysis when the relay signal-to-noise ratio (SNR) is fixed $\gamma_R=3$ dB, and the destination SNR changes $\gamma_D \in \{0,1,\dots,6\}$ dB. 
        The lines represent the mutual information obtained with the learned CF strategy (marginal scheme in Fig.~\ref{fig:marg_model}), as a function of the destination SNR $\gamma_D$.
        The dotted lines represent models trained for a single value of $\gamma_D=\gamma_R=\gamma$.
        The solid blue line represents the model trained over equal SNR at both the relay and the destination $\gamma_D=\gamma_R =\gamma \sim \text{Unif.}\{0,1,\dots,6\}$ dB.
        The green triangles represent the model trained for a fixed SNR at the relay $\gamma_R=3$ dB, and variable SNR at the destination $\gamma_D \sim \text{Unif.}\{0,1,\dots,6\}$ dB. 
    }
    \label{fig:robustness-fixSNRrelay}
\end{figure}

Fig.~\ref{fig:robustness-fixSNRrelay} shows the mutual information achieved as a function of the SNR at the destination $\gamma_D\in\{0,1,\dots,6\}$ dB, when the SNR at the relay is fixed as $\gamma_R=3$ dB.
In other words, the statistics of the relay's received signal do not change, while the received signal $Y_D$ at the destination has variable SNR levels.
We also include two robust models, one trained for a single $\gamma_R=3$ dB, and a range of $\gamma_D\sim\text{Unif.}\{0,1,\dots,6\}$ dB, and another trained for SNRs $\gamma_D=\gamma_R=\gamma \sim\text{Unif.}\{0,1,\dots,6\}$ dB.
We note that the performance of the model trained on a range (with the same SNR on both $\gamma_D=\gamma_R=\gamma$) is equivalent to the performance of the robust model trained for [$\gamma_R=3$ dB, $\gamma_D\in\{0,1,\dots,6\}$ dB].

\subsubsection{Testing on a range of SNRs at the relay, fixing the SNR at the destination}

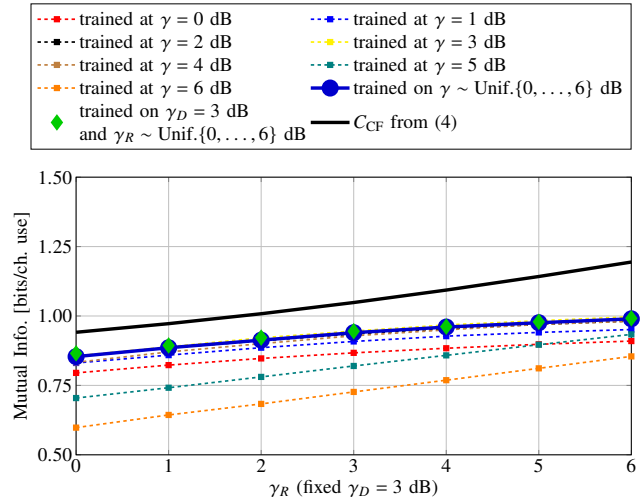
\begin{figure}
    \centering
    \pgfplotsset{
    compat=newest,
    /pgfplots/legend image code/.code={%
        \draw[mark repeat=2,mark phase=2,#1] 
            plot coordinates {
                (0cm,0cm) 
                (0.6cm,0cm)
                (1.2cm,0cm)%
            };
    },
}

\begin{tikzpicture}
\begin{axis}[
	scale only axis,
	font=\LARGE,
	width=\textwidth,
	height=0.5*\textwidth,
	xmajorgrids=true,
	ymajorgrids=true,
	xmin=0,
	xmax=6,
        xtick={0,1,2,3,...,10},
	ymin=0.5,
	ymax=1.5,
        y tick label style={
        /pgf/number format/.cd,
            fixed,
            fixed zerofill,
            precision=2,
        /tikz/.cd
    },
    ytick={0,0.25,...,10},
	xlabel={$\gamma_R$ (fixed $\gamma_D=3$ dB)},
        ylabel={Mutual Info. [bits/ch. use]},
	legend columns=2,
	legend cell align={left},
	legend style={row sep=3pt, at={(1,1.1)}, anchor = south east, inner sep =5pt, align=left},
    cycle list name=color list,
	]

    \addplot+[line width=1.5pt, dashed, mark=square*, mark options={scale=1,fill,solid}] table[x=X, y=Y, col sep=tab] 
    {
X	Y
0	0.7950072289
1	0.8230773807
2	0.8471393585
3	0.8672283292
4	0.8838329315
5	0.8979722261
6	0.909678638
    };

    \addplot+[line width=1.5pt, dashed, mark=square*, mark options={scale=1,fill,solid}] table[x=X, y=Y, col sep=tab] 
    {
X	Y
0	0.8307488561
1	0.8590043187
2	0.8858116269
3	0.9082942009
4	0.9273654222
5	0.9408195615
6	0.9511858225
    };

    \addplot+[line width=1.5pt, dashed, mark=square*, mark options={scale=1,fill,solid}] table[x=X, y=Y, col sep=tab] 
    {
X	Y
0	0.8543393016
1	0.8846924901
2	0.9105564952
3	0.9352453947
4	0.9542581439
5	0.9695264697
6	0.9820777774
    };

    \addplot+[line width=1.5pt, dashed, mark=square*, mark options={scale=1,fill,solid}] table[x=X, y=Y, col sep=tab] 
    {
X	Y
0	0.8563659787
1	0.8890722394
2	0.9214341044
3	0.9457474947
4	0.9677583575
5	0.9828975797
6	0.9972350001
    };

    \addplot+[line width=1.5pt, dashed, mark=square*, mark options={scale=1,fill,solid}] table[x=X, y=Y, col sep=tab] 
    {
X	Y
0	0.8331068754
1	0.8704930544
2	0.9004092216
3	0.9286572337
4	0.9499971271
5	0.9696590304
6	0.980938077
    };

    \addplot+[line width=1.5pt, dashed, mark=square*, mark options={scale=1,fill,solid}] table[x=X, y=Y, col sep=tab] 
    {
X	Y
0	0.7043981552
1	0.7415660024
2	0.7805520892
3	0.8199461102
4	0.8582172394
5	0.8966156244
6	0.9336596727
    };

    \addplot+[line width=1.5pt, dashed, mark=square*, mark options={scale=1,fill,solid}] table[x=X, y=Y, col sep=tab] 
    {
X	Y
0	0.5979005694
1	0.643550694
2	0.6833053231
3	0.7264849544
4	0.7687656283
5	0.8114676476
6	0.8544861078
    };

    \addplot+[line width=3pt, mark=*, \myBlue, solid, mark options={scale=3,fill=\myBlue,solid}] table[x=X, y=Y, col sep=tab] 
    {
X	Y
0	0.853681922
1	0.8855768442
2	0.9130349159
3	0.9398543239
4	0.9601229429
5	0.9757241011
6	0.9897454381
    };

    \addplot+[line width=4pt, green!80!black, solid, mark=diamond*, only marks, mark options={scale=2,fill=green!80!black,solid}] table[x=X, y=Y, col sep=tab] 
    {
X	Y
0	0.8655198812
1	0.8939297795
2	0.9220933914
3	0.9452043772
4	0.9615089297
5	0.9791213274
6	0.9916504025
    };

    \addplot+[line width=3pt, mark=none, black, solid] table[x=X, y=Y, col sep=tab] 
    {
X	Y
0	0.9413192947
1	0.9723704524
2	1.00808198
3	1.04848204
4	1.093356518
5	1.142216239
6	1.194287945
    };

    \legend{
        trained at $\gamma=0$ dB,
        trained at $\gamma=1$ dB,
        trained at $\gamma=2$ dB,
        trained at $\gamma=3$ dB,
        trained at $\gamma=4$ dB,
        trained at $\gamma=5$ dB,
        trained at $\gamma=6$ dB,
        trained on {$\gamma\sim \text{Unif.}\{0,\dots,6\}$} dB,
        trained on {$\gamma_D=3$ dB} \\ {and $\gamma_R\sim \text{Unif.}\{0,\dots,6\}$} dB,
        $C_\text{CF}$ from~\eqref{eq:C_CF-Simeone}
    }
    
\end{axis}

\end{tikzpicture}
    \caption{
        Robustness analysis when the destination signal-to-noise ratio (SNR) is fixed $\gamma_D=3$ dB, and the relay SNR changes $\gamma_R \in \{0,1,\dots,6\}$ dB. 
        The lines represent the mutual information obtained with the learned CF strategy (marginal scheme in Fig.~\ref{fig:marg_model}), as a function of the relay SNR $\gamma_R$.
        The dotted lines represent models trained for a single value of $\gamma_D=\gamma_R=\gamma$.
        The solid blue line represents the model trained over equal SNR at both the relay and the destination $\gamma_D=\gamma_R=\gamma \sim \text{Unif.}\{0,1,\dots,6\}$ dB.
        The green diamonds represent the model trained for a fixed SNR at the destination $\gamma_D=3$ dB, and variable SNR at the destination $\gamma_R \sim \text{Unif.}\{0,1,\dots,6\}$ dB. 
    }
    \label{fig:robustness-fixSNRdest}
\end{figure}

Fig.~\ref{fig:robustness-fixSNRdest} illustrates the mutual information as a function of the SNR at the relay $\gamma_R\in\{0,1,\dots,6\}$ dB, when the SNR at the destination is fixed as $\gamma_D=3$ dB.
In this case, the received signal $Y_D$ at the destination has fixed statistics, while the relay's received signal is subjected to different SNR levels.  As above, we include two robust models, one trained for a single $\gamma_D=3$ dB, and a range of $\gamma_R\sim\text{Unif.}\{0,1,\dots,6\}$ dB, and another one trained for SNRs $\gamma_D=\gamma_R=\gamma \sim\text{Unif.}\{0,1,\dots,6\}$ dB.
Similar to the previous scenario, the performance of the model trained on a range (with the same SNR on both $\gamma_D=\gamma_R=\gamma$) is equivalent to the performance of the robust model trained for [$\gamma_D=3$ dB, $\gamma_R\in\{0,1,\dots,6\}$ dB]. 
We also note that, in this scenario, the baseline models trained at an SNR in the vicinity of $\gamma_D=\gamma_R=\gamma=3$ dB perform well.  

In summary, we showed that training on a range of {\em equal} SNRs for both at the relay and the destination provides a good compromise in performance. This suggests good generalization capabilities for both the compressor and the demodulator, eliminating the need for ad-hoc SNR choices during training.   
Experimental results suggest that knowing the SNR at the destination is generally more important in order to achieve good performance.
In principle, the destination could have a fine-tuned model for each SNR (or SNR range) it experiences. Concurrently, the experiments demonstrate that training robust relay nodes only requires a rough estimate of the SNR range at the relay.

\section{Conclusion and Future Work}
\label{sec:conclusion}
In this paper, we have revisited CF relaying in the context of learned distributed compression and incorporated a task-oriented neural WZ compressor into a PRC setup as a practical form of CF relaying mechanism.  
Our proposed framework represents the first proof-of-concept work for an interpretable learned CF relaying scheme, where both the compressor and the demodulator components are parameterized with lightweight ANNs. 
Such a design choice also enables us to provide post-hoc explanations of these learned components by explicitly visualizing their behaviors.
Our results demonstrate that the learned CF schemes exhibit characteristics of the optimal asymptotic CF, such as binning of the quantized indices at the relay. 
We also note that the performance of these schemes, across various modulation schemes (both real and complex-valued), meets the communication rate of perfect relay ($R\to\infty$) with minimal relay rate $R$.
We have also demonstrated that training over a range of SNRs,  both at the destination and the relay, provides good generalization over the range of interest, with minimal performance degradation compared to models trained for a specific SNR.

Extending our framework to a general relay channel, in which the destination does successive decoding of the compressed relay index and the source information, would be possible.  Additional design constraints arising from incorporating a learned CF in full-duplex and half-duplex relay channels, as well as more complex and realistic channel models would be interesting future research directions. Another promising area for future exploration is extending the proposed neural CF frameworks to handle multi-hop networks and MIMO relay channels.

\bibliographystyle{./bibliography-1571030310/IEEEtran}

\bibliography{./bibliography-1571030310/references}

\clearpage

\begin{IEEEbiographynophoto}{Ezgi~{\"O}zyılkan} is a Ph.D. candidate at New York University's Tandon School of Engineering, advised by Professor Elza Erkip. She obtained her M.Eng. degree from Imperial College London in 2021 under the supervision of Professor Deniz Gündüz. Her research focuses on learning-based, interpretable approaches for data compression and communication problems, particularly in distributed scenarios. Ezgi has served as a reviewer for both machine learning and information theory publications, such as IEEE Transactions on Information Theory, IEEE Transactions on Communications, International Symposium on Information Theory (ISIT), and International Conference on Machine Learning (ICML). Ezgi has co-organized several workshops, including the Machine Learning and Compression Workshop at NeurIPS 2024 and ``Learn to Compress" at ISIT 2024. Ezgi's research has been recognized with numerous awards, including the Ivor Tupper Prize from Imperial College, the Future Leader Ph.D. Fellowship from NYU Tandon, and the IEEE Signal Processing Society Scholarship. She is a member of IEEE.
\end{IEEEbiographynophoto}

\begin{IEEEbiographynophoto}{Fabrizio Carpi}
received his Ph.D. in Electrical and Computer Engineering from New York University (NYU) in 2024, where he was also a member of the NYU WIRELESS research center. 
He received his M.S. in Communication Engineering from the University of Parma, Italy, in 2018. 
He received the Best Poster Award at the IEEE Communication Theory Workshop (CTW) in 2021, and the Best Student Paper Award at the IEEE Internation Workshop on Signal Processing Advances in Wireless Communications (SPAWC) in 2021. 
His research interests include communication and information theory, wireless communications, task-aware source coding, and applied machine learning.
\end{IEEEbiographynophoto}

\begin{IEEEbiographynophoto}{Siddharth Garg} received the B.Tech. degree in electrical engineering from the Indian Institute of Technology Madras, Chennai, India, and the Ph.D. degree in electrical and computer engineering from Carnegie Mellon University, Pittsburgh, PA, in 2009. 
He is currently an Associate Professor with New York University, New York, where he joined as an assistant professor in 2014. Prior to this, he was an Assistant Professor with the University of Waterloo, Waterloo, ON, Canada, from 2010 to 2014. 
His current research interests include computer engineering, and more particularly in secure, reliable, and energy efficient computing. 
He was a recipient of the NSF Career Award in 2015, and the paper awards at the IEEE Symposium on Security and Privacy 2016, the USENIX Security Symposium in 2013, the Semiconductor Research Consortium TECHCON in 2010, and the International Symposium on Quality in Electronic Design in 2009. 
He was listed in popular science magazine’s annual list of “Brilliant 10” researchers. 
He serves on the technical program committee of several top conferences in the area of computer engineering and computer hardware and has served as a reviewer for several IEEE and ACM journals conferences.
\end{IEEEbiographynophoto}

\begin{IEEEbiographynophoto}{Elza Erkip}
is an Institute Professor in the Electrical and Computer Engineering Department at New York University Tandon School of Engineering. She received the B.S. degree in Electrical and Electronics Engineering from Middle East Technical University, Ankara, Turkey, and the M.S. and Ph.D. degrees in Electrical Engineering from Stanford University, Stanford, CA, USA.  Her research interests are in information theory, communication theory, and wireless communications.

Dr. Erkip is a member of the Science Academy of Turkey and is a Fellow of the IEEE. She received the NSF CAREER award in 2001, the IEEE Communications Society WICE Outstanding Achievement Award in 2016, the IEEE Communications Society Communication Theory Technical Committee (CTTC) Technical Achievement Award in 2018, and the IEEE Communications Society Edwin Howard Armstrong Achievement Award in 2021. She was the Padovani Lecturer of the IEEE Information Theory Society in 2022. Her paper awards include the IEEE Communications Society Stephen O. Rice Paper Prize in 2004,  the IEEE Communications Society Award for Advances in Communication in 2013 and  the IEEE Communications Society Best Tutorial Paper Award in 2019.  She was a member of the Board of Governors of the IEEE Information Theory Society 2012-2020, where she was the President in 2018.   She was a Distinguished Lecturer of the IEEE Information Theory Society from 2013 to 2014. She is currently the Editor-in-Chief of IEEE Journal on Selected Areas in Information Theory and the Chair of IEEE Communications Society Communication Theory Technical Committee.

\end{IEEEbiographynophoto}

\end{document}